\PassOptionsToPackage{pdfpagelabels=false}{hyperref} 
\documentclass[fleqn,usenatbib,useAMS]{mnras}
\usepackage{xspace}

\usepackage{hyperref}

\usepackage{graphicx}
\usepackage{inputenc}
\usepackage[T1]{fontenc}
\usepackage{lmodern}
\usepackage{inputenc}
\usepackage{epsfig}
\usepackage{hyperref}
\usepackage{amsmath}
\usepackage{lmodern}
\usepackage[T1]{fontenc}
\usepackage{natbib}
\usepackage{relsize}
\usepackage{ae,aecompl}
\usepackage{newtxtext,newtxmath}
\usepackage{graphicx}	
\usepackage{amsmath}	
\usepackage{amssymb}	

\newcommand{\kms}{km~s$^{-1}$ }
\newcommand{\kmsp}{km~s$^{ -1}$}
\newcommand{\oi}{\ion{O}{i}~}

\newcommand{\siii}{\ion{Si}{ii}~}

\newcommand{\sii}{\ion{S}{ii}~}
\newcommand{\siiifs}{\ion{Si}{ii}$^\ast$~}
\newcommand{\siiv}{\ion{Si}{iv}~}

\newcommand{\oip}{\ion{O}{i}}

\newcommand{\ovi}{\ion{O}{vi}~}

\newcommand{\siiip}{\ion{Si}{ii}}

\newcommand{\siiiip}{\ion{Si}{iii}}

\newcommand{\siip}{\ion{S}{ii}}

\newcommand{\siivp}{\ion{Si}{iv}}

\newcommand{\galex}{{\it GALEX} }

\newcommand{\mout}{$\dot{M}_{\text{o}}$ }
\newcommand{\mstar}{$M_\ast$ }
\newcommand{\mstarp}{$M_\ast$}

\newcommand{\moutp}{$\dot{M}_{\text{o}}$}
\newcommand{\eoutp}{$\dot{E}_{\text{o}}$}

\newcommand{\esfrp}{$\dot{E}_{\text{SFR}}$}
\newcommand{\esfr}{$\dot{E}_{\text{SFR}}$ }
\newcommand{\poutp}{$\dot{p}_{\text{o}}$}
\newcommand{\pout}{$\dot{p}_{\text{o}}$ }
\newcommand{\psfrp}{$\dot{p}_{\text{SFR}}$}
\newcommand{\psfr}{$\dot{p}_{\text{SFR}}$ }

\newcommand{\vcircp}{v$_{\text{circ}}$}
\newcommand{\vcirc}{v$_{\text{circ}}$ }

\defcitealias{chisholm15}{Paper I}	
\defcitealias{chisholm16}{Paper II}
\defcitealias{chisholm16b}{Paper {\small III}}

\begin{document}

\title[Mass and Momentum Outflow Rates]{The Mass and Momentum Outflow Rates of Photoionized Galactic Outflows}
\author[Chisholm et al.]{John Chisholm$^{1}$\thanks{Contact email: John.Chisholm@unige.ch}, Christy A. Tremonti$^{2}$, Claus Leitherer $^{3}$, Yanmei Chen$^{4}$\\
$^{1}$  Observatoire de Gen\`{e}ve, Universit\`{e} de Gen\`{e}ve, 51 Ch. des Maillettes, 1290 Versoix, Switzerland \\
$^{2}$Astronomy Department, University of Wisconsin, Madison, 475
  N. Charter St., WI 53711, USA \\
$^{3}$Space Telescope Science Institute, 3700 San Martin Drive, Baltimore, MD 21218, USA \\
$^{4}$Department of Astronomy, Nanjing University, Nanjing 210093, China\\
}
\pubyear{2016}
\label{firstpage}
\pagerange{\pageref{firstpage}--\pageref{lastpage}}
\maketitle
\begin{abstract}
Galactic outflows are believed to play an important role in regulating star formation in galaxies, but estimates of the outflowing mass and momentum have historically been based on uncertain assumptions. Here, we measure the mass, momentum, and energy outflow rates of seven nearby star-forming galaxies using ultraviolet absorption lines and observationally motivated estimates for the density, metallicity, and radius of the outflow. Low-mass galaxies generate outflows faster than their escape velocities with mass outflow rates up to twenty times larger than their star formation rates. These outflows from low-mass galaxies also have momenta larger than provided from supernovae alone, indicating that multiple momentum sources drive these outflows. Only 1-20\% of the supernovae energy is converted into kinetic energy, and this fraction decreases with increasing stellar mass such that low-mass galaxies drive more efficient outflows. We find scaling relations between the outflows and the stellar mass of their host galaxies (M$_\ast$) at the 2-3$\sigma$ significance level. The mass-loading factor, or the mass outflow rate divided by the star formation rate, scales as M$_\ast^{-0.4}$ and with the circular velocity as v$_\mathrm{circ}^{-1.6}$. The scaling of the mass-loading factor is similar to recent simulations, but the observations are a factor of five smaller, possibly indicating that there is a substantial amount of unprobed gas in a different ionization phase. The outflow momenta are consistent with a model where star formation drives the outflow while gravity counteracts this acceleration.
\end{abstract}

\section{INTRODUCTION}
In actively star forming galaxies, high-mass stars inject energy and momentum into the surrounding gas, heating and accelerating the gas out of star-forming regions as a galactic outflow \citep{heckman2000, veilleux05, erb15}. Models suggest that  removing this residual mass and energy from star-forming regions regulates star formation in galaxies \citep{dekel86, white, katz96, hopkins14}. Consequently, galactic outflows are a ubiquitous component of modern galaxy formation and evolution simulations \citep{springel03, hopkins14, voglesberger14,  schaye15}. However, it is computationally infeasible, at the moment, to fully resolve all of the necessary physics to drive galactic outflows, and simulations typically scale the mass and energy outflow rates with properties of their host galaxy, like the star formation rate (SFR) or stellar mass  \citep[\mstarp;][]{springel03, oppenheimer06, somerville15}.

The mass and energy outflow rates are challenging to observe. Outflows are diffuse structures with uncertain geometries that span different ionization states and metallicities \citep{heckman90, heckman2000, veilleux05}, which are challenging to observationally constrain. Many studies use a single absorption line to trace both the density and kinematic information of the outflowing gas, but rely on assumptions for the ionization corrections, metallicities, and outflow radii to derive mass outflow rates \citep{rupkee2005, martin2005, weiner, rubin13, heckman15}. These assumptions may introduce an order of magnitude scatter into the observed relations \citep{murray07}.

\begin{table*}
\begin{tabular}{lcccccccccc}
\hline
Galaxy name & log(\mstarp) & \vcircp   & SFR & SFR$_\mathrm{COS}$ & log(O/H)+12 & Proposal ID & References  \\
 & [log(M$_\odot$)]  & [\kmsp] & [M$_\odot$~yr$^{-1}$]& [M$_\odot$~yr$^{-1}$]& [dex]& &  \\
\hline
SBS~1415+437 &6.9 & 18& 0.02 & 0.02 & 7.6 & 11579 & 8 \\
1~Zw~18 &7.2 & 21 &  0.02 & 0.02 & 7.2 & 11579 & 8 \\
MRK~1486 &9.3 & 82 &3.6 & 2.5&  8.1 & 12310 &  2, 6, 10, 11, 12 \\
KISSR~1578 & 9.5 & 94 & 3.7 & 2.1 & 8.1 & 11522 & 3, 14 \\
Haro~11 & 10.1 & 137 & 26 & 12 & 8.1 &13017 & 1, 7   \\
NGC~7714 &10.3 & 156 & 9.2 & 3.1 & 8.5 & 12604 & 4, 5, 13\\
IRAS~08339+6517 & 10.5 & 179 & 14 & 4.7 & 8.5 &  12173 & 9\\
NGC~6090 & 10.7 & 202& 25& 5.5 & 8.8& 12173 & 9\\
\end{tabular}
\caption{Derived galaxy properties (see \autoref{samp} for sample selection details). The first column gives the name of the galaxy, the second column gives the stellar mass (\mstarp), the third column gives the circular velocity of the galaxy (\vcircp) calculated using a Tully-Fischer relation \citep{reyes}, the forth column gives the star formation rate of the entire galaxy (SFR), the fifth column corrects the total SFR for the fact that COS resolves a portion of the total galaxy (SFR$_\mathrm{COS}$), the sixth column is the oxygen abundances (log(O/H)+12) taken from \citet{chisholm15}, the seventh column is the HST proposal ID, and the last column lists previous references for the data that we use. References for the targets are coded as: (1) \citet{alexandroff}, (2) \citet{duval}, (3) \citet{france2010}, (4) \citet{fox2013}, (5) \citet{fox14}, (6) \citet{hayes14}, (7) \citet{heckman15}, (8) \citet{james}, (9) \citet{claus2012}, (10) \citet{ostlin}, (11) \citet{pardy}, (12) \citet{Rivera}, (13) \citet{richter2013}, and (14) \citet{wofford2013}.}
\label{tab:sample}
\end{table*}

\begin{table*}
\begin{tabular}{lcccccccccc}
\hline
Galaxy name & \moutp/SFR$_\mathrm{COS}$ &  \eoutp/\esfrp & \poutp/\psfrp & M$_\mathrm{o}$ & R$_\mathrm{i}$  \\
& & & & (10$^6$~M$_\odot$)& (pc)  \\
\hline
SBS~1415+437 & $19 \pm 17$ & $0.21 \pm 0.18$ & $5.7 \pm 5.0$ & 0.2 & 40 \\
1~Zw~18 & $11 \pm 8.0$ & $0.15 \pm 0.11$ & $3.7 \pm 2.6$  & 0.3 & 62\\
MRK~1486 & $0.91 \pm 0.33$ & $0.03 \pm 0.01$ & $0.48 \pm 0.18$ & 1 & 45 \\
KISSR~1578 & $2.0 \pm 0.57$ & $0.09 \pm 0.03$ & $1.27 \pm 0.35$ & 2 &  54\\
Haro~11 &$1.2 \pm 0.51$ & $0.05 \pm 0.02$ & $0.70 \pm 0.29$ & 14 & 101 \\
NGC~7714 & $0.23 \pm 0.08$ & $0.009 \pm 0.003$ & $0.13 \pm 0.04$ & 0.34 & 33 \\
IRAS~08339+6517 & $0.07 \pm 0.03$ & $0.007 \pm 0.004$ & $0.06 \pm 0.03$ & 0.02 & 13   \\
NGC~6090 & $0.54 \pm 0.18$ & $0.02 \pm 0.006$ & $0.28 \pm 0.09$ & 1.9 & 63 \\
\end{tabular}
\caption{Derived outflow properties. The first column gives the name of the galaxy, the second column column gives the maximum mass outflow rate divided by the SFR in the COS aperture (\moutp/SFR$_\mathrm{COS}$), the third column the maximum energy outflow rate divided by the energy from supernovae within the COS aperture (\eoutp/\esfrp), the fourth column gives the maximum momentum outflow rate divided by the direct momentum from supernovae within the COS aperture (\poutp/\psfrp), the fifth column gives the mass at the inner radius of the outflow, and the last column gives the inner radius of the outflow. We calculate the errors using a Monte Carlo method (see \autoref{masses} for details). All the values for IRAS~08339+6517 are upper limits because the profile is not fit by our model, and we exclude this galaxy from the sample (see \autoref{iras}).}
\label{tab:sample_out}
\end{table*}

In a series of papers we have used a sample of nearby star-forming galaxies with ultraviolet spectra from the Cosmic Origins Spectrograph \citep[COS;][]{cos} on the {\it Hubble Space Telescope} to characterize the physical conditions within galactic outflows. In \citet{chisholm15} (hereafter \citetalias{chisholm15}), we define a sample of 48 nearby star-forming galaxies, our method to fit the stellar continuum, and characterize the absorption kinematics. We then use the \siii absorption profiles to find shallow scaling relations between outflow velocities and their host galaxy properties. In \citet{chisholm16} (hereafter \citetalias{chisholm16}) we extend this analysis to the \oip, \siiiip, and \siiv transitions, which span a factor of three in ionization potential, and probe both neutral and ionized gas. We find that the moderately ionized \siiv and neutral \oi are co-moving, implying that we are observing a single outflowing structure. We then use the equivalent width ratios to study the ionization structure of the outflows, and find that the outflow equivalent widths are reproduced by photoionization models, if the observed O and B stars ionize the outflow. Using these photoionization models, we estimate the ionization structure, the metallicity, and the total hydrogen density at the base of the outflow. These values vary from galaxy-to-galaxy, and each galaxy requires a unique photoionization model. Finally, in \citet{chisholm16b} (hereafter \citetalias{chisholm16b}) we fit detailed models of the \siivp~1402\AA\ line profile to determine the acceleration, the radial density structure, and the inner radius of the outflow from the starburst NGC~6090. We combine these measurements with detailed photoionization models to derive a velocity resolved mass outflow rate with observationally motivated values for the metallicity, ionization correction, and physical extent of the outflow.

Here, we extend the analysis of \citetalias{chisholm16b} to a sample of 7 nearby star-forming galaxies with the highest signal-to-noise UV spectra. We first briefly describe the data analysis and how we characterize the outflow (\autoref{data}). In \autoref{scale}, we then examine the physical picture suggested by the observations, and use it to inform our calculation of the mass, momentum, and energy outflow rates (\autoref{masses}). In \autoref{global} we normalize the outflow energetics by the mass and energy produced by supernovae, and explore the scaling relations with the stellar mass of the galaxies. Finally, in \autoref{discussion} we compare the values to previous observations and simulations (\autoref{comp}), explore the implications of these relations for driving galactic outflows (\autoref{driving}), and discuss a galaxy that our model does not fit (\autoref{iras}). In a companion paper, we will explore how the physical properties of galactic outflows (metallicity, density, radius and velocity structure) scale with host galaxy properties, and their implications for the mass-metallicity relation and the enrichment of the circum-galactic medium.

\section{DATA AND ANALYSIS}
\label{data}

The data reduction and methods follow \citetalias{chisholm15} and \citetalias{chisholm16b}. Here we summarize the major steps taken, but refer the reader to those papers for details. We first discuss the sample (\autoref{samp}), the data reduction, and the continuum fitting (\autoref{cont}). We then discuss how we fit the line profiles of the \siivp~1402\AA\ doublet to derive important velocity-resolved relations (\autoref{profile}). Finally, in \autoref{photo} we describe how we use \oip~1302\AA, \siiip~1304\AA, \siip~1250\AA, and \siivp~1402\AA\ column densities, along with {\small CLOUDY} photoionization models, to derive the metallicities, densities, and ionization corrections of the outflows.

\subsection{Sample selection}
\label{samp}
We select the eight galaxies from \citetalias{chisholm15} that have COS spectra with a signal-to-noise ratio greater than five at 1380~\AA\ (near the \siiv line; the median signal-to-noise ratio of the sample is 11); a measured central velocity less than 0 at the 1$\sigma$ significance for the \oi~1302, \siii~1304, \sii~1250, {\it and} \siiv~1402~\AA\ lines; and are not contaminated by geocoronal emission at both the \oi~1302 and \siii~1304~\AA\ absorption lines. The most crucial, and constraining, requirement is that we cleanly observe all of the transitions because multiple transitions with different ionization potentials constrain the ionization structure and metallicity of the outflow. These cuts produce a sample of eight high-quality spectra that are kinematically defined as outflows. A list of the previous {\it Hubble Space Telescope} proposal identifications and references for each galaxy is given in \autoref{tab:sample}.

Since COS is a circular aperture spectrograph, the spectral resolution varies depending on the size of the target, with the resolution of our sample varying from 21~\kms to 58~\kmsp, as measured from the Milky Way absorption lines \citepalias{chisholm15}. In \citetalias{chisholm15} we calculate the star formation rates (SFR) and stellar masses (\mstarp) of the host galaxies, assuming a  Chabrier initial mass function \citep{chabrier}, using archival {\it WISE} \citep{wise} and {\it GALEX} \citep{galex} observations \citep{buat11, jarrett2013, querejeta}. These values are recorded in \autoref{tab:sample}. While the sample is small, it represents the only high signal-to-noise spectra in the HST archive for which we can accurately calculate the mass outflow rates. 

\subsection{Observations and continuum fitting}
\label{cont}
The COS spectra are processed through the {\small CalCOS} pipeline, version 2.20.1, and downloaded from the MAST server. The individual exposures are combined and wavelength calibrated following the methods outlined in \citet{wakker2015}. The spectra are normalized, binned by 5 pixels (a spacing of 10~\kmsp), and smoothed by 3 pixels.

A linear combination of multi-age, fully theoretical {\small STARBURST99} models \citep{claus99, claus2010}, computed using the WM-Basic stellar library and the Geneva stellar evolution tracks with high-mass loss \citep{geneva94}, are fit to the spectra following the methods of \citetalias{chisholm15}. We simultaneously fit for the reddening along the line-of-sight using a Calzetti extinction law \citep{calzetti}. We use these fits to remove contributions from the stellar continuum, set the zero-velocity of the absorption lines, and to describe the number of ionizing photons in the photoionization models below. 

\subsection{Line profile fitting}
\label{profile}
For each galaxy we simultaneously fit the velocity-resolved \siiv optical depth ($\tau$) and covering fraction (C$_f$) to describe the variation of the line profile with velocity. The radiative transfer equation defines the flux ($F$) at a given velocity in terms of the C$_f$ and the $\tau$ at that velocity as
\begin{equation}
 F(\mathrm{v}) = 1-C_f(\mathrm{v}) +C_f(\mathrm{v})e^{-\tau(\mathrm{v})}
\end{equation}
If there is a doublet, where the transitions share the same C$_f$ and the ratio of the two optical depths is equal to the ratio of their $f$-values, then a system of equations solves for both C$_f$ and $\tau$ independently \citep{hamann}. We use the \siiv doublet to form a system of equations and solve for the velocity-resolved C$_f$ and $\tau$ in terms of the flux of the \siiv doublet as
\begin{equation}
 \begin{aligned} 
    C_f(\mathrm{v}) &= \frac{\mathrm{F}_\mathrm{W}(\mathrm{v})^2-2\mathrm{F}_\mathrm{W}(\mathrm{v}) + 1}{\mathrm{F}_\mathrm{S}(\mathrm{v})-2\mathrm{F}_\mathrm{W}(\mathrm{v})+1}\\
    \tau(\mathrm{v}) &= \ln\left(\frac{C_f(\mathrm{v})}{C_f(\mathrm{v})+F_\mathrm{W}(\mathrm{v})-1}\right)
\end{aligned}
\label{eq:cf}
\end{equation}
Where F$_\mathrm{W}$ is the {\small STARBURST99} continuum normalized flux of the weaker doublet line (\siiv 1402\AA), and F$_\mathrm{S}$ is the continuum normalized flux of the stronger doublet line (\siiv 1394\AA). The $\tau$ and C$_f$ errors are calculated by varying the observed fluxes by a Gaussian kernel centered on zero with a standard deviation equal to the error on the flux measurement. We then recalculate the $\tau$ and C$_f$ values at each velocity interval, and repeat the procedure 1000 times to form $\tau$ and C$_f$ distributions. We take the standard deviation of these distributions as the velocity-resolved $\tau$ and C$_f$ errors. To increase the signal-to-noise ratio, we further bin these profiles by two pixels. The $\tau$ and C$_f$ distributions for each galaxy are shown in \autoref{appendix}. Now, we fit physically motivated, velocity-resolved models to the observed $\tau$ and C$_f$ distributions.

To fit the $\tau$ profile we assume that the \siiv density ($n_\mathrm{4}$) follows a radial power-law such that
\begin{equation}
n_\mathrm{4}\left(r\right) = n_\mathrm{4, 0} \left(\frac{r}{\mathrm{R}_\mathrm{i}}\right)^{\alpha}
\label{eq:denlaw}
\end{equation}
where $n_\mathrm{4, 0}$ is the \siiv density at the base of the outflow, R$_\mathrm{i}$ is the initial radius of the outflow, and $\alpha$ is an unknown power-law exponent. As discussed in \citetalias{chisholm16b}, a power-law distribution with $\alpha = -2$ is expected if the outflow is mass-conserving. In this scenario, the density falls as 1/r$^2$  as the mass in a thin shell is spread over a successively larger surface area as it moves out in the flow. This radial decline may be due to either deviations in geometry from spherically symmetric outflow geometries or because  mass is removed from the flow as it propagates outward.  In the local galaxy M~82, \citet{leroy15} find a steep density power-law, with an $\alpha$ between $-3$ and $-5$, implying that the outflow is not a mass-conserving flow. Therefore, \autoref{eq:denlaw} is general, and allows for either a mass-conserving outflow ($\alpha=-2$) or one that does not conserve mass ($\alpha < -2$).

Since C$_f$ measures the fraction of the continuum area covered by the absorbing gas, C$_f$ will change as the solid angle of the absorbing clouds change \citep{martin09, steidel10}. We fit the covering fraction with a radial power-law such that
\begin{equation}
C_f (r) = C_f(\mathrm{R}_\mathrm{i}) \left(\frac{r}{\mathrm{R}_\mathrm{i}}\right)^{\gamma}
\end{equation}
where $C_f(\mathrm{R}_\mathrm{i})$ is the covering fraction of the outflow at R$_\mathrm{i}$,and $\gamma$ is the power-law exponent that measures the decline of C$_f$ with radius. As gas moves away from the continuum source, the total area at a given radius increases. If the clouds remain the same size, this geometric dilution will cause C$_f$ to fall as r$^{-2}$. However, if the clouds expand adiabatically some of the geometric dilution can be offset, and C$_f$ declines more gradually. In \citetalias{chisholm16b}, we measure a $\gamma$ of -0.8 from the \siiv profile of NGC~6090, consistent with expectations of adiabatically expanding clouds in an adiabatically expanding medium. Moreover, \citet{steidel10} fit the C$_f$ profiles of $z \approx 2$ Lyman Break Galaxies and find $\gamma$ values between $-0.2$ and $-0.6$.

The radial velocity structure of these lines are unknown, a priori, and we assume that the velocity follows a $\beta$-law \citep{lamers} such that
\begin{equation}
\mathrm{v} (r) = \mathrm{v}_\infty \left(1-\frac{\mathrm{R}_\mathrm{i}}{r}\right)^{\beta}
\label{eq:beta}
\end{equation}
where v$_\infty$ is the measured terminal velocity of the outflow. A $\beta$-law is commonly used to describe stellar winds \citep{lamers}. Analytic relations from \citet{murray05} suggest that driving outflows with optically thin radiation pressure or ram pressure lead to a $\beta$ near $-0.5$, consistent with the \siiv profile from NGC~6090 \citetalias{chisholm16b}.  We simplify Equations 3-5 by introducing the normalized quantities $w$ = v/v$_\infty$ and $x$ = $r/$R$_\mathrm{i}$ to give the relation for the normalized velocity as $w = (1-1/x)^\beta$.

As found in \citetalias{chisholm16b}, the equivalent width ratios of the \siiv 1394\AA\ and \siiv 1402\AA\ lines indicate that the \siiv is not heavily saturated. Therefore, we simultaneously fit the observed velocity resolved optical depth and covering fraction, assuming a Sobolev optical depth \citep{sobolev, lamers, prochaska2011, scarlata}, as
\begin{equation}
\begin{aligned}
\tau\left(w\right) &= \frac{\upi e^2}{\mathrm{mc}} f \lambda_\mathrm{0} \frac{\mathrm{R}_\mathrm{i}}{\mathrm{v}_\infty} \mathrm{n}_\mathrm{4, 0} x^\alpha \frac{dx}{dw}= \tau_0 \frac{w^{1/\beta-1}}{\beta(1-w^{1/\beta})^{2+\alpha}}\\
C_f(w) &=  \frac{C_f (\mathrm{R}_\mathrm{i})}{\left(1-w^{1/\beta}\right)^\gamma}
\label{eq:cfbeta}
\end{aligned}
\end{equation}
where m is the mass of the electron, $f$ is the oscillator strength of the \siiv 1402\AA\ line, $\lambda_0$ is the restframe wavelength of the \siiv 1402\AA\ line, and $\tau_0$ is a constant term that we define as
\begin{equation}
\tau_0 = \frac{\upi e^2}{\mathrm{mc}} f \lambda_\mathrm{0} \frac{\mathrm{R}_\mathrm{i}}{\mathrm{v}_\infty} \mathrm{n}_\mathrm{H,0} \chi_\mathrm{Si 4} \mathrm{(Si/H)}
\label{eq:tau0}
\end{equation}
where we have replaced n$_\mathrm{4,0}$ with the hydrogen density at the base of the outflow (n$_\mathrm{H,0}$), the \siiv ionization fraction of the outflow ($\chi_\mathrm{Si 4}$), and the silicon to hydrogen abundance of the outflow (Si/H). These parameters are estimated from the observed column densities and {\small CLOUDY} models in \autoref{photo} below.

We fit \autoref{eq:cfbeta} to the velocity resolved $\tau$ and C$_f$ distributions of the \siiv 1402\AA\ line using {\small MPFIT} \citep{mpfit} to solve for $\tau_0$, $\beta$, $\alpha$, $\gamma$, C$_f$(R$_\mathrm{i}$). Resonant emission decreases the C$_f$ and increases the $\tau$ of the outflow \citep[][]{prochaska2011}, leading to unphysical velocity distributions if the resonance emission is not accounted for (see gray points in \autoref{appendix}). As in \citetalias{chisholm16b}, we use the \siiifs 1194\AA\ non-resonant emission line as a proxy of resonance emission, and mask-out velocities with observed \siii 1194\AA\ emission. The \siiifs emission lines are narrow, and typically only occupy the inner $\pm50$~\kms of the profile. Therefore, low velocity absorption--which can also be contaminated with ISM absorption from within the host galaxy--is largely excluded from the fit. The fitted relations describe the velocity structure of the outflowing \siiv gas, and the profile fits for the full sample are described fully in a future paper but are shown in \autoref{appendix}. While most galaxies have similar fit parameters (median plus standard deviation for $\beta$, $\gamma$, $\alpha$ are $0.5\pm 0.1$, $-0.9 \pm 0.7$, $-5.3 \pm 2.4$ respectively), one galaxy, IRAS~08339+6517, has a C$_f$ distribution which cannot be fit by \autoref{eq:cfbeta} (see \autoref{fig:iras}). Instead we force a flat unity C$_f$ distribution ($\gamma$ = 0 and  C$_f$(R$_\mathrm{i}$) = 1).  Since IRAS~08339+6517 does not follow \autoref{eq:cfbeta}, we exclude it from our outflow discussion, but include an upper limit of its mass outflow rate using $\gamma = 0$ and C$_f$(R$_\mathrm{i}$)~=~1 on all plots as an X. In \autoref{iras} we discuss this profile further.

\subsection{Photoionization modeling}
\label{photo}

The column densities of the different transitions trace the ionization structure of the outflow. To characterize this ionization structure, we measure column densities (N) of the \oip~1302\AA, \siiip~1304\AA, \siip~1250\AA, and \siivp~1402~\AA\ lines. The measured lines are weak: the \siiv doublet equivalent width ratio is greater than 1.3 for all galaxies (median of 1.7), while the $f$-value ratio is 2.0, indicating that the \siiv 1402\AA\ line is not overly saturated. We measure column densities assuming that the lines are partially covered, as measured by $C_f (\mathrm{R}_\mathrm{i})$ in \autoref{eq:cfbeta}. Partial coverage is important, especially for low-mass galaxies with low covering fractions; not accounting for partial coverage artificially decreases N. We determine N by integrating the optical depth over a by-eye determined velocity range using the following expression
\begin{equation}
    N = \frac{3.77 \times 10^{14}~\text{cm}^{-2}}{\lambda[\text{\AA}] f} \int \mathrm{ln}\left(\frac{C_f(\mathrm{R}_\mathrm{i})}{C_f(\mathrm{R}_\mathrm{i})+F(\mathrm{v})-1}\right) \mathrm{dv}
\end{equation}
where $F($v$)$ is the {\small STARBURST99} stellar continuum normalized flux. We assume that C$_f$(R$_\mathrm{i}$) remains the same for each of the four transitions, consistent with the result from \citetalias{chisholm16} that the line width, not the covering fraction, changes from transition to transition. 

We fit the observed column densities to integrated column densities from a large grid of {\small CLOUDY}, version 13.03, photoionization models \citep{ferland}, which use the observed {\small STARBURST99} stellar continuum fit as the ionizing source. Additionally, we use an expanding spherical geometry with the measured density profiles from \autoref{profile}.  We create the {\small CLOUDY} models by varying the ionization parameter, metallicity, and density of the outflow; and tabulate the column densities for each model. We use {\small CLOUDY}'s \ion{H}{ii} abundances, including the Orion dust grain distribution which accounts for depletion of metals onto grains, from \citet{baldwin1991}. We scale these abundances between 0.01~Z$_\odot$ and 2.5~Z$_\odot$ to create the grid of {\small CLOUDY} models. We then infer the best-fit ionization parameters, metallicities, and densities using a Bayesian approach with the \oip, \siiip, \siip, and \siiv column densities from the observations and the {\small CLOUDY} models  \citepalias{chisholm16b}. From these models we estimate $\chi_{\siivp}$, the outflow metallicity (Z$_\mathrm{o}$), and n$_\mathrm{H,0}$. The ionization fractions and metallicities convert n$_\mathrm{H, 0}$ to the \siiv density at the base of the outflow (n$_\mathrm{4,0})$, which, with the fitted \siiv optical depth and maximum \siiv outflow velocity, defines R$_\mathrm{i}$ \citepalias[see \autoref{eq:tau0};][]{chisholm16b}. We can now calculate the masses and energetics of the outflows.

\section{MASSES AND ENERGETICS OF OUTFLOWS}
\label{mout}
Here we calculate the mass outflow rates and energetics of the galactic outflows. In \autoref{energy} we consider what drives the outflows: energy and momentum from star formation. By comparing the outflow mass, momentum and energy to the corresponding quantities released through star formation, we study how efficiently outflows remove mass and momentum from galaxies. We then consider the physical situation of galactic outflows by asking: on which physical scales do we observe outflows (\autoref{scale})? The answer to this question affects how we interpret the energetics of the outflows. Finally, in \autoref{masses} we calculate the mass outflow rates (\moutp), energy outflow rates (\eoutp), and momentum outflow rates (\poutp); and derive how these quantities scale with the stellar mass of the galaxy (\autoref{global}).

\subsection{Energetics of star formation}
\label{energy}
Star formation deposits energy and momentum into gas, accelerating it out of star-forming regions \citep{mckee, dekel86, murray05}. Therefore, to know the efficiency of a galactic outflow one must know the amount of energy deposited into the surrounding gas by star formation. For example, comparing \mout to the mass of stars formed per unit time (SFR) describes how efficiently galaxies convert gas into stars. The ratio of \mout to SFR is called the mass-loading factor. Similarly, we normalize the momentum and energy of the outflow by the momentum and energy released by star formation. \citet{claus99} give the energy deposition rate from supernovae as
\begin{equation}
\dot{E}_\mathrm{SFR} = 3 \times 10^{41} \left(\frac{\mathrm{SFR}}{1 M_\odot ~\mathrm{yr}^{-1}}\right)~\mathrm{g~cm}^2~\mathrm{s}^{-3}
\label{eq:esfr}
\end{equation}
 and \citet{murray05} give the direct momentum deposition from supernovae as
 \begin{equation}
\dot{p}_\mathrm{SFR} = 2 \times~10^{33} \left(\frac{\mathrm{SFR}}{1 M_\odot ~\mathrm{yr}^{-1}}\right)~\mathrm{g~cm~s}^{-2}
\label{eq:psfr}
\end{equation} 
Where both of these relations assume one supernovae per 100~M$_\odot$ of stars formed. These relations only include the direct energy and momentum from supernovae; other energy and momentum sources may change these relations. For \psfrp,  radiation pressure, the Sedov Taylor phase, and stellar winds from massive stars inject at least as much momentum as supernovae \citep{claus99, kim15}. Below, we normalize the mass, momentum and energy outflow rates by their star formation quantities to study how efficiently outflows remove mass, momentum and energy from star-forming regions. 

\subsection{Which physical scales are outflows observed?}
\label{scale}

Our observations indicate that the covering fraction declines with radius (C$_f \propto $ r$^{-0.9}$) and the observed gas density drops dramatically with radius  (n~$\propto$~r$^{-5.3}$). Further, the measured initial radii of the outflows indicate that they begin nearly at the size of the star-forming regions.  However, this does not mean that there is not absorption from larger radii along the line-of-sight. The equivalent widths of our outflows are typically 6-10 times larger than the equivalent widths seen from studies of the circum-galactic medium \citep{steidel10, werk13}. This means that the circum-galactic absorption is likely blended within the outflow absorption at low, zero, or positive velocities, contaminating trends of outflow centroid velocities, as discussed in \citetalias{chisholm15}. While warm gas has been observed out to 100~kpc in the circum-galactic medium \citep{steidel10, tumlinson, werk}, a majority of the gas we observe at high velocities (< -200~\kmsp) likely comes from within 300~pc of the star-forming region \citepalias{chisholm16b}. 

Recently, \citet{thompson16} proposed a plausible scenario that links outflows at small radii to circum-galactic absorption at large radii. High-energy photons, stellar winds, and supernovae inject energy and momentum into the surrounding gas, thermalizing some gas into a hot, 10$^7$~K, outflow and accelerating the residual gas out of the star-forming region \citep{weaver, chevalier, heckman2000, cooper08, bustard}. The cooler photoionized gas observed here is produced either during the snowplow phase of the hot supernovae ejecta \citep{mckee, maclow}, or it is ambient gas entrained in the hot outflow \citep{cooper09}. Either way, the photoionized gas is radially accelerated, destroyed and incorporated into the hotter outflow on short timescales \citep{klein, scannapieco15, bruggen}. This process "mass-loads" the wind, increasing the mass of the hot outflow \citep{strickland09}, and accounts for the rapid drop in \siiv density \citepalias[see \autoref{profile};][]{chisholm16b}. The hot mass-loaded phase continues out of the disk, where it encounters and interacts with the hot ejecta from other localized star-forming regions, producing the shocked shell-like structures and filaments seen in recombination lines in M~82 \citep{shopbell}. This hotter outflow continues out of the galaxy, undetected by the photoionized tracers here, adiabatically and radiatively cooling as it travels out. Eventually the hot outflow cools to temperatures near the peak of the cooling curve where it shocks, reproducing photoionized gas at large radii \citep{thompson16}. 

This physical picture means that the observed outflows are powered by the local star formation within the COS aperture, and the outflow properties depend on the local star formation \citep{bordoloi16}. Consequently, the driving source of the outflow is not the global SFR, rather the local SFR. Our global SFRs are calculated from a combination of the FUV and FIR luminosities \citepalias{chisholm15}. We cannot use this method for the local SFRs because we lack high-resolution FUV and FIR imaging.  However, we can measure the FUV in our aperture directly from the COS spectra. Assuming that the fraction of unobscured-to-total SFR is the same in the COS aperture as in the galaxy as a whole we calculate:  
\begin{equation}
 \mathrm{SFR}_\mathrm{COS} = \mathrm{SFR} \times \frac{F_\mathrm{COS}}{F_\mathrm{GALEX}}
\end{equation}
Where SFR is the total SFR calculated using the IR and UV luminosities, $F_\mathrm{COS}$ is the flux in the COS aperture, and F$_\mathrm{GALEX}$ is the \galex measured flux. The lowest mass galaxies have similar SFR$_\mathrm{COS}$ to their total SFR because the COS aperture encloses all of the {\it GALEX} flux, whereas the higher mass galaxies have spatially extended star formation outside of the COS aperture. This model assumes that the IR light follows the UV light, whereas the SFR of high-mass galaxies may be clumpy. F$_\mathrm{COS}$/F$_\mathrm{GALEX}$ ranges from 1, for low-mass galaxies, to 0.22, for the highest mass galaxies. The SFR$_\mathrm{COS}$ values for the full sample are given in \autoref{tab:sample}.  

\subsection{Mass and energy outflow rates}
\label{masses}
Here we estimate the mass, energy, and momentum outflow rates. In \citetalias{chisholm16b}, we calculate the mass outflow rate (\moutp) at each velocity interval using the profile fitting and the photoionization modelling as 
\begin{equation}
\begin{aligned}
    \dot{M}_\mathrm{o}(\mathrm{r}) &= \Omega C_{f}(\mathrm{r}) \mathrm{v}(\mathrm{r}) \rho(\mathrm{r}) \mathrm{r}^2 \\
    \dot{M}_\mathrm{o}(w)&= \Omega C_f(\mathrm{R}_\mathrm{i}) \mathrm{v}_\infty \mu \mathrm{m}_\mathrm{p}  \mathrm{n}_\mathrm{H,0} \mathrm{R}_\mathrm{i}^2 \frac{w}{(1-w^{1/\beta})^{2+\gamma+\alpha}}
    \label{eq:mout}
\end{aligned}
\end{equation}
Where $\mu\mathrm{m}_\mathrm{p}$ is is the average mass per nucleon, or 1.4 times the proton mass for standard abundances \citep{asplund}; $w$ is the velocity normalized by v$_\infty$; and $\Omega$ is the solid angle occupied by the outflow, which we assume is 4$\upi$ because a locally driven outflow is not yet collimated by the disk. The relations for C$_f$(r) and $\rho$(r) are taken from the profile fitting (\autoref{eq:cf} and \autoref{eq:denlaw}), and the values for n$_\mathrm{H,0}$ and R$_\mathrm{i}$ are taken from the photoionization modeling (\autoref{photo}) and \autoref{eq:tau0}, respectively. The total mass at each velocity interval is similarly calculated as
\begin{equation}
    M_\mathrm{o} (w) = \Omega C_f(\mathrm{R}_\mathrm{i}) \mu \mathrm{m}_\mathrm{p} \mathrm{n}_\mathrm{H,0} \mathrm{R}_\mathrm{i}^3 \left(\frac{1}{1-w^{1/\beta}}\right)^{3+\alpha+\gamma}
\end{equation}
We then calculate the energy outflow rate (\eoutp) using \autoref{eq:mout} as
\begin{equation}
\dot{E}_\mathrm{o}(w) = \frac{1}{2} \dot{M}_\mathrm{o}(w) v_\infty^2 w^2
\end{equation}
and the momentum outflow rate (\poutp) as
\begin{equation}
\dot{p}_\mathrm{o} (w) = \dot{M}_\mathrm{o}(w) v_\infty w
\end{equation}
These four quantities are velocity resolved. Specifically, \mout increases at low-velocities and decreases at high-velocities, reaching a maximum \mout at intermediate velocities. The increase in \mout happens as the velocity and radius increase, and the decrease happens as the density and covering fraction decrease \citepalias{chisholm16b}. Here, we take the maximum value of each quantity  as the estimate of the quantity. This means that the reported values in \autoref{tab:sample} and in each figure are calculated at specific velocities that correspond to their maximum values. We choose the maximum value as the representative value because a radially accelerated outflow implies that the \mout in each velocity interval is a snapshot of the \mout at a given velocity (or equivalently radius or time).   Further, if the decrease in density is due to a phase change (photoionized gas to a hotter phase), then the decrease in \mout actually represents a transfer of \mout from the photoionized phase to a hotter phase. In this case, the maximum \mout represents the time when the photoionized outflow is the largest contributor to the total \mout of the galaxy.

The errors of each quantity are calculated by varying the estimated parameters of \autoref{eq:mout} by a Gaussian distribution centered on zero with a standard deviation corresponding to the parameters' measured errors. We then recalculate the \mout value with these Monte Carloed values, and repeat the procedure 1000 times to form a \mout distribution. We take the standard deviation of this distribution as the errors on \mout and propagate the errors accordingly for \pout and \eoutp. The errors are larger for low-mass galaxies because narrow absorption line profiles are challenging to determine the density scaling ($\alpha$), which leads to larger uncertainties in $\alpha$ and \moutp. We normalize each of the quantities by the SFR, star formation energy deposition rate (\autoref{eq:esfr}) and star formation momentum deposition rate (\autoref{eq:psfr}) within the COS aperture to determine how efficiently outflows remove these quantities from star-forming regions (see \autoref{tab:sample_out} for the values).
\subsubsection{Scaling relations}
\label{global}

\begin{figure}
\includegraphics[width = 0.5\textwidth]{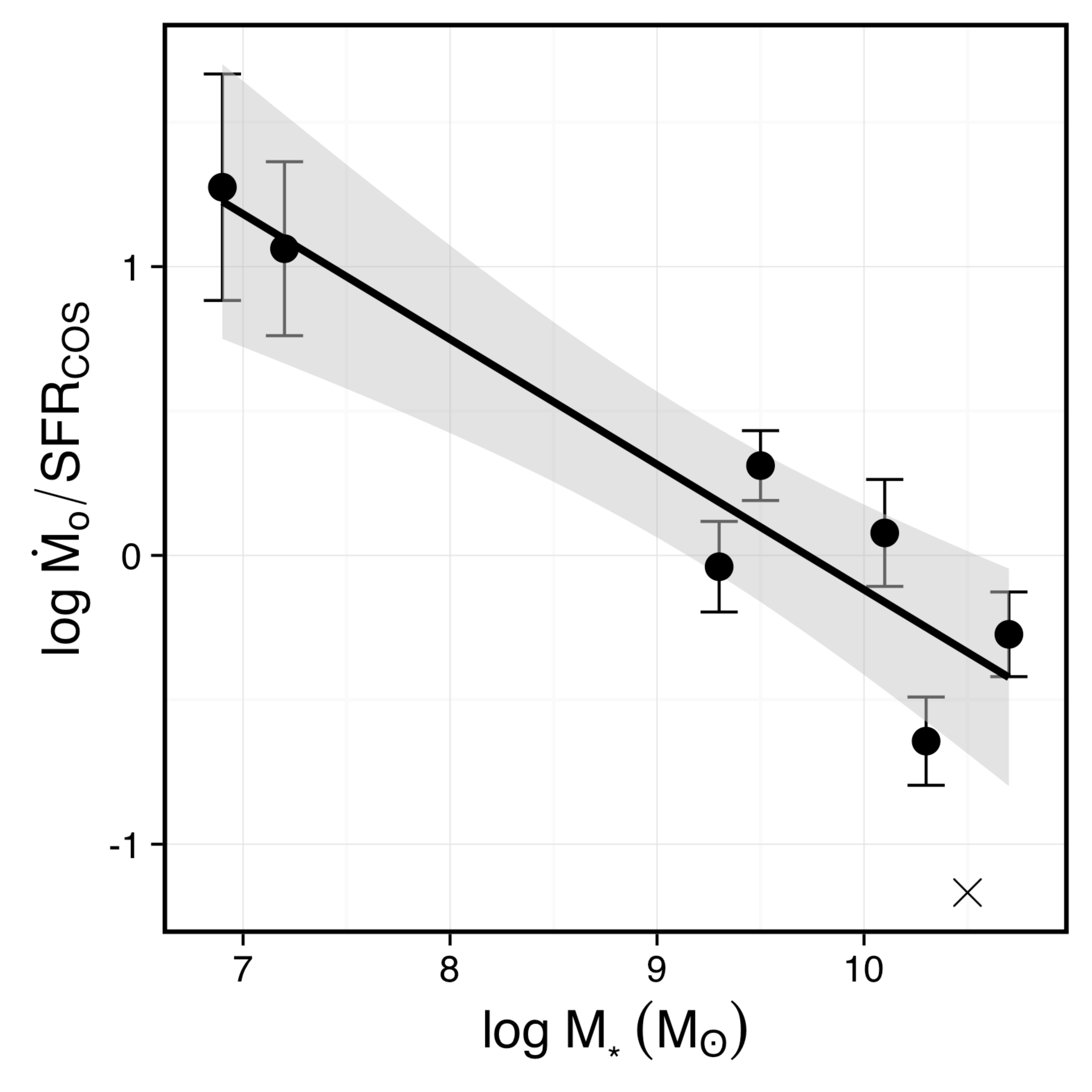}
\caption{The scaling of the maximum mass-loading factor ($\dot{M}_\mathrm{o}$/SFR$_\mathrm{COS}$) with stellar mass (\mstarp). The line gives the least squares regression fit to the circles (see \autoref{eq:eta}), while the gray region is the 95\% confidence interval of the fit. The X is IRAS~08339+6517, a high-mass merger that is not fit by the line profile model of \autoref{eq:cfbeta} (see \autoref{iras}). }
\label{fig:mout}
\end{figure}

Here, we study how the masses and energetics of outflows scale with the stellar mass of their host galaxies. With only a sample of seven galaxies, more high-quality data is required to confirm these relations, but we do find statistically significant correlations. As discussed in \autoref{iras}, we exclude IRAS08339 from the fits because the line profile does not follow the model of \autoref{profile}, and upper-limits of the \mout estimates are shown on the plots as an X, assuming $\gamma=0$ and C$_f$(R$_\mathrm{i})=1$. \autoref{fig:mout} shows the scaling of the mass-loading factor with stellar mass. Over-plotted in black is the least-squares fit to the relation, with the 95\% confidence interval as the gray region. This trend corresponds to a relation of
\begin{equation}
\frac{\dot{M}_\mathrm{o}}{\mathrm{SFR_\mathrm{COS}}} = 0.76 \pm 0.20 \left(\frac{\mathrm{M}_\ast}{10^{10}~\mathrm{M}_\odot}\right)^{-0.43 \pm 0.07}
\label{eq:eta}
\end{equation}
The fit is significant at the 3$\sigma$ significance level (p-value < 0.001), and has a coefficient of determination (R$^2$) of 0.88, where an R$^2$ of 1.0 implies that the fit describes 100\% of the variation. The relation has a residual standard error of 0.26~dex.

\begin{figure}
\includegraphics[width = 0.5\textwidth]{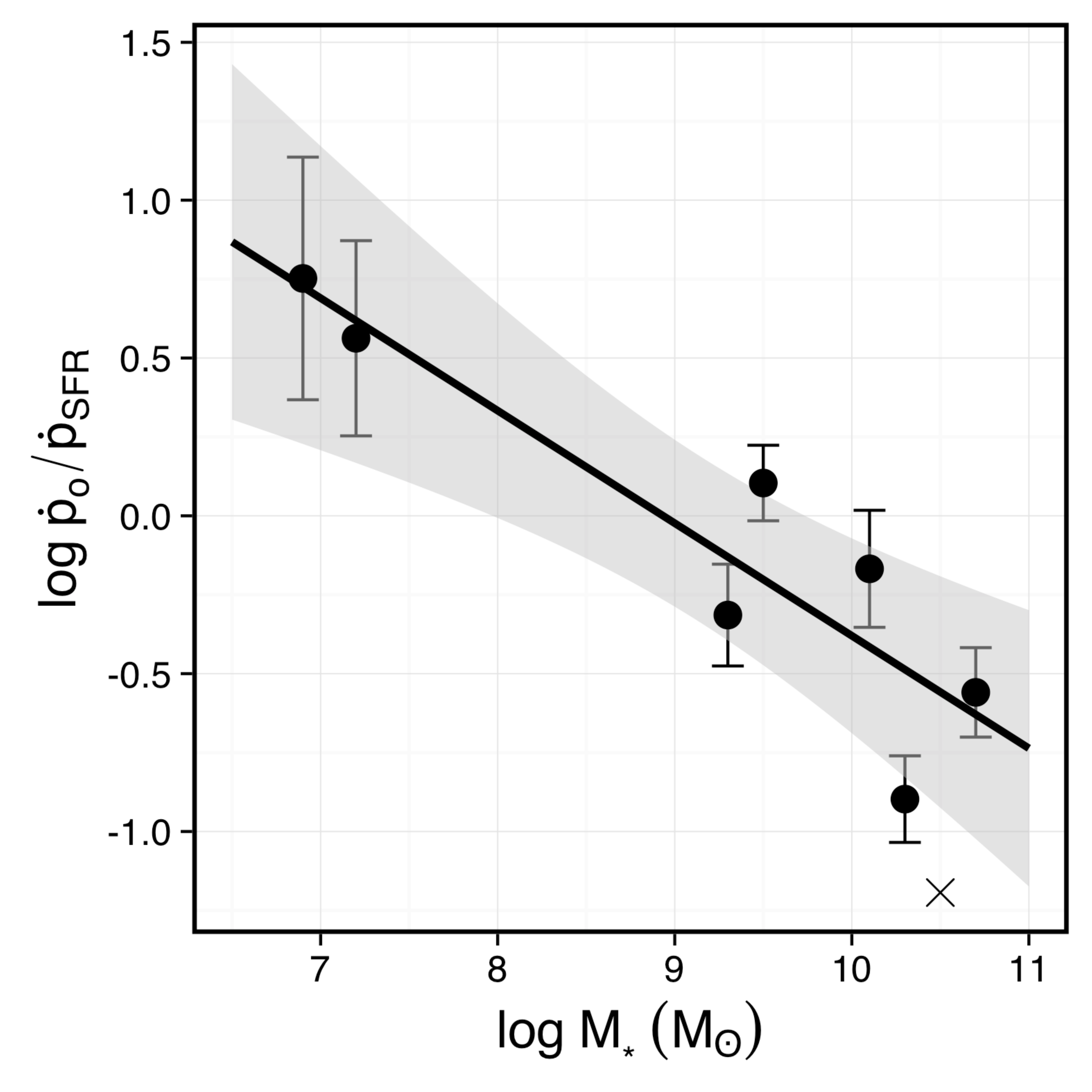}
\caption{The outflow momentum efficiency ($\dot{p}_\mathrm{o}$/\psfrp) with stellar mass (\mstarp). The line gives the least squares regression fit to the circles (see \autoref{eq:pout}), while the gray region is the 95\% confidence interval of the fit. The X is IRAS~08339+6517, a high-mass merger that is not fit by the line profile model of \autoref{eq:cfbeta} (see \autoref{iras}).}
\label{fig:pout}
\end{figure}

\begin{figure}
\includegraphics[width = 0.5\textwidth]{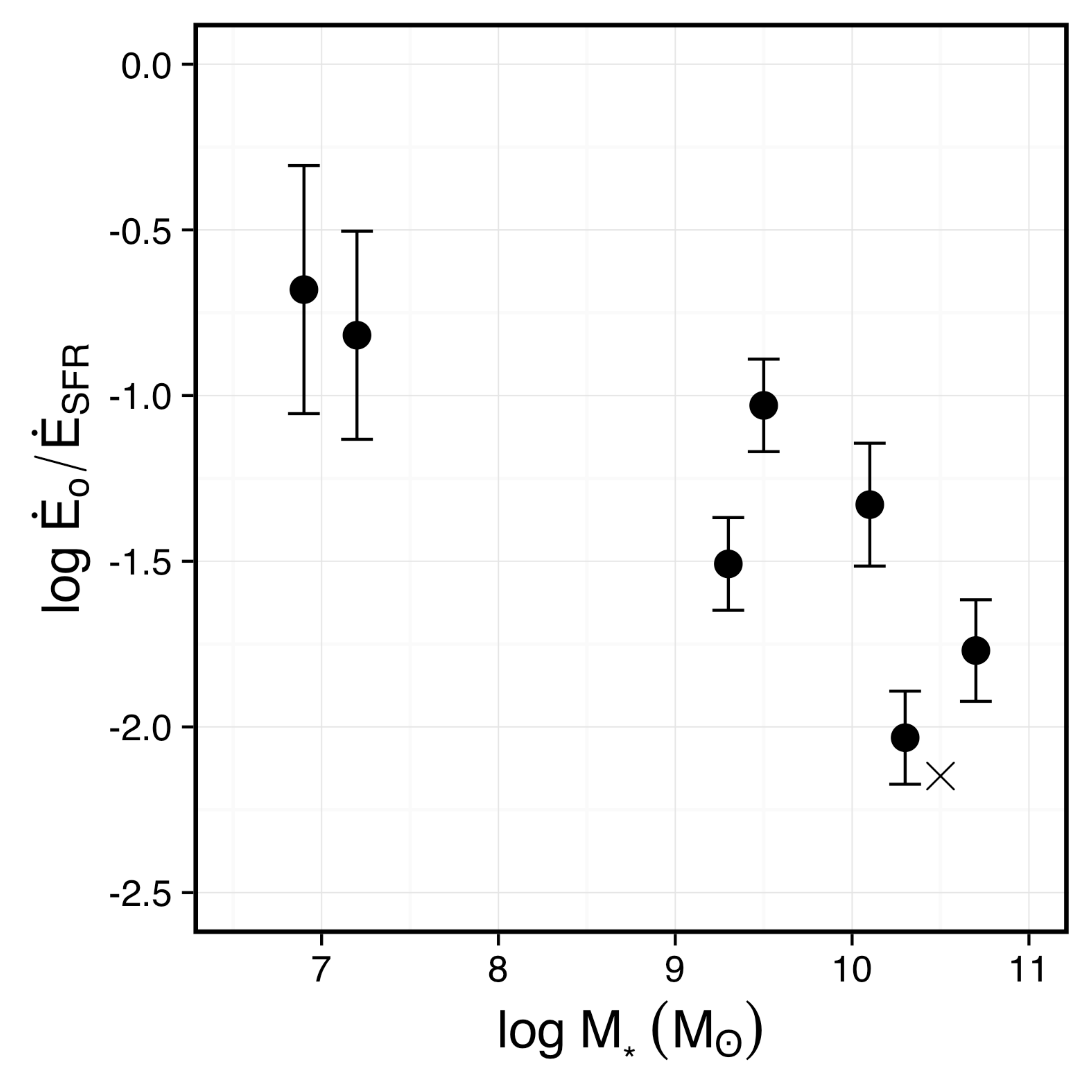}
\caption{The outflow energy efficiency  ($\dot{E}_\mathrm{o}$/\esfrp) with stellar mass  (\mstarp). We do not find a significant correlation between $\dot{E}$$_\mathrm{o}$/\esfr and \mstarp, although there is a trend such that higher mass galaxies have lower $\dot{E}$$_\mathrm{o}$/\esfrp. The X is IRAS~08339+6517, a high-mass merger that is not fit by the line profile model of \autoref{eq:cfbeta} (see \autoref{iras}).}
\label{fig:eout}
\end{figure}

Several simulations scale the mass-loading factor by the circular velocity (\vcircp) of the galaxy \citep[e.g.,][]{somerville15}. Since we do not measure \vcirc for our sample, we rescale \mstar into \vcirc using the Tully-Fisher relation from \citet{reyes}.  Doing this, we find
\begin{equation}
\frac{\dot{M}_\mathrm{o}}{\mathrm{SFR_\mathrm{COS}}} = 1.12 \pm 0.27 \left(\frac{\mathrm{v}_\mathrm{circ}}{100~\mathrm{km/s}}\right)^{-1.56 \pm 0.25}
\label{eq:etavcirc}
\end{equation}
Similarly, \autoref{fig:pout} gives the scaling of the momentum efficiency with \mstar as
\begin{equation} 
\begin{aligned}
\frac{\dot{p}_\mathrm{o}}{\dot{p}_\mathrm{SFR}} &= 0.42 \pm 0.12 \left(\frac{\mathrm{M}_\ast}{10^{10}~\mathrm{M}_\odot}\right)^{-0.36 \pm 0.07}\\
&= 0.58 \pm 0.15 \left(\frac{\mathrm{v}_\mathrm{circ}}{100~\mathrm{km/s}}\right)^{-1.39 \pm 0.26}
\label{eq:pout}
\end{aligned}
\end{equation}
which is significant at the 2.5$\sigma$ level (p-value < 0.006), and has an R$^2$ of 0.85.  Finally, in \autoref{fig:eout} we show the variation of the outflow energy efficiency with \mstarp, which has a scaling relation of
\begin{equation}
\begin{aligned}
\frac{\dot{E}_\mathrm{o}}{\dot{E}_\mathrm{SFR}} &= 0.03 \pm 0.01 \left(\frac{\mathrm{M}_\ast}{10^{10}~\mathrm{M}_\odot}\right)^{-0.28 \pm 0.07}\\
&= 0.04 \pm 0.01 \left(\frac{\mathrm{v}_\mathrm{circ}}{100~\mathrm{km/s}}\right)^{-1.02 \pm 0.27}\\
\label{eq:eout}
\end{aligned}
\end{equation}
This relation is only significant at the 2$\sigma$ level (p-value < 0.01), and has an R$^2$ of 0.74. We do not consider this relation significant, but an inverse relation exists such that low-mass galaxies have higher energy efficiencies (the trend has a Kendall's $\tau$ coefficient of $-0.71$). These relations describe how efficiently photoionized outflows remove mass, momentum and energy from star-forming regions. 

\section{DISCUSSION}
\label{discussion}
\subsection{Comparison to previous observations and simulations}
\label{comp}
The most important results of this study are the values of the mass outflow rate, the momentum outflow rate, and the energy outflow rate (\autoref{tab:sample_out}) that are estimated using observationally motivated values for the outflow  metallicity, ionization fraction, and radius. Additionally, we use this small sample to find significant (>2.5$\sigma$) scaling relations between the mass (\autoref{fig:mout}) and momentum (\autoref{fig:pout}) of the photoionized outflows and the stellar mass of their host galaxies (\mstarp).  

The mass-loading values found here are broadly consistent with the wide range of mass-loading factors found in the literature \citep{rupkee2005, weiner, rubin13}. Three of the galaxies studied here are included in the \citet{heckman15} sample, which calculates \mout assuming a constant outflow column density of 10$^{20.85}$~cm$^{-2}$, a constant outflow metallicity of 0.5~Z$_\odot$, and that the outflow radius is a constant factor of two times the radius of the star-forming region. Our estimates of the mass-loading factors for these three galaxies are, on average, a factor of three times smaller than theirs,  however, the difference has a large range: between an over-estimate of 1.04 and 5.7 times. This discrepancy is largely due to our observationally motivated radii, metallicities, and ionization corrections \citepalias{chisholm16b}, which vary by factors of 3, 10 and 3, respectively, for the sample. \citet{heckman15} find a statistically weak correlation between \moutp/SFR and \vcirc of  \moutp/SFR~=~1.8~(v$_\mathrm{circ}$/100~\kmsp)$^{-0.98}$. This relation has a normalization that is 1.6 times larger and a shallower scaling than \autoref{eq:etavcirc}. Crucially, by not accounting for galaxy-to-galaxy variations in the metallicities, ionization corrections and radii, previous studies may introduce up to a factor of 10 scatter into the  \mout relations, possibly obscuring trends. 

\begin{figure}
\includegraphics[width = 0.5\textwidth]{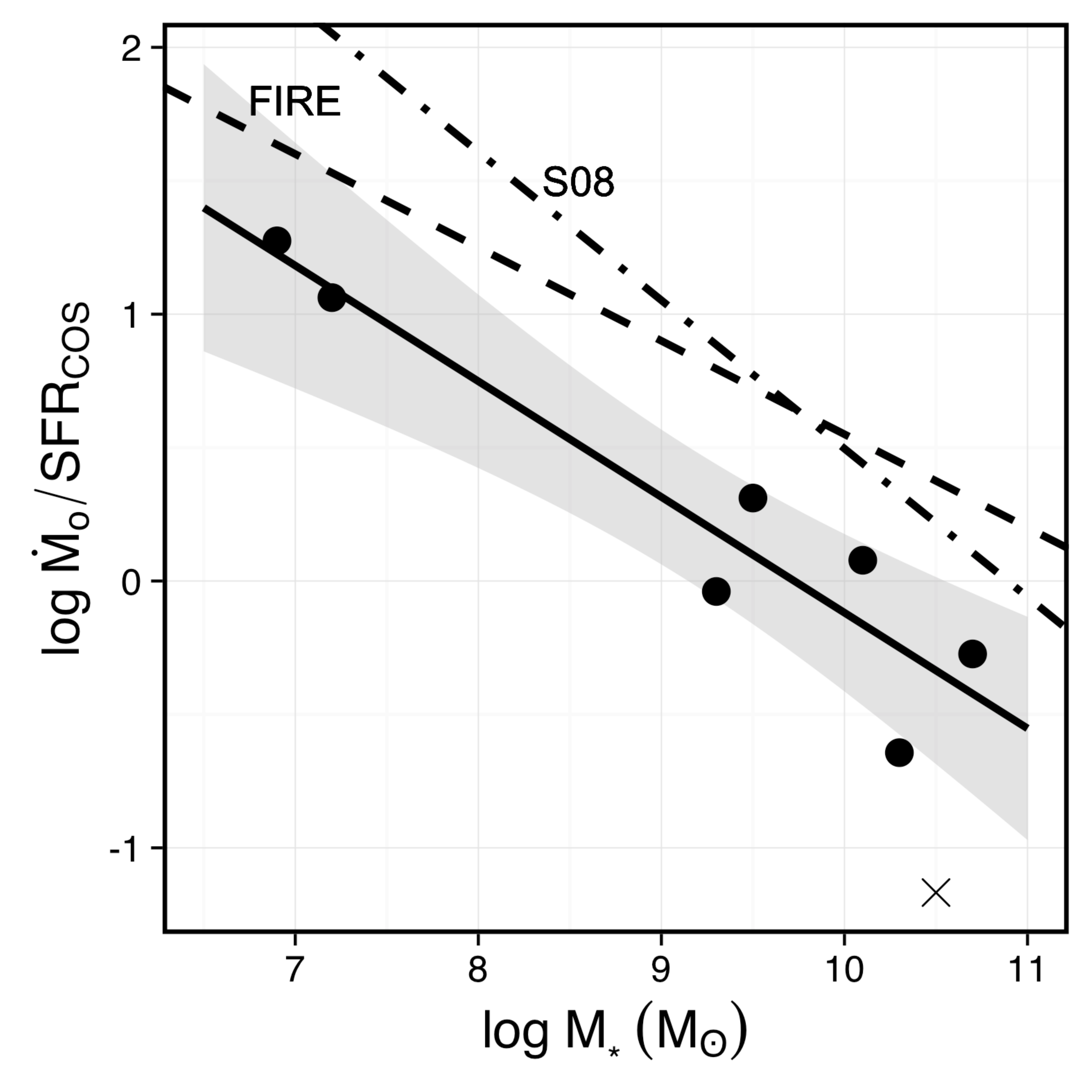}
\caption{Comparison of the observed mass-loading factors (\moutp/SFR$_\mathrm{COS}$) with typical relations from simulations. The solid line gives the least squares regression fit to the circles (see \autoref{eq:eta}), while the gray region is the 95\% confidence interval of the fit. The dashed line shows the best-fit relation from the FIRE simulations \citep[calculated at 0.25R$_\mathrm{vir}$;][]{muratov} and the dot-dashed line shows the relation from \citet{somerville} (S08). The relations from simulations are typically a factor of five larger than the observed photoionized outflows. The X is IRAS~08339+6517, a high-mass merger with a line profile that is not fit by our model (see \autoref{iras}).}
\label{fig:moutann}
\end{figure}

We also compare these results to relations typically used in simulations. Simulations use a variety of scaling relations to drive galactic outflows \citep{somerville15}, including scaling the mass-loading factor as v$_\mathrm{circ}^{-1}$ \citep{oppenheimer08, dutton} or as v$_\mathrm{circ}^{-2}$ \citep{benson03, somerville}. In \autoref{fig:moutann} we over-plot the \citet{somerville} relation on our \mout estimates as a dot-dashed line. This relation is steeper and has a larger normalization than the observations presented here.

Recent high-resolutions simulations produce outflows without explicitly scaling the outflow properties to the host galaxy. One example of this is the FIRE simulation \citep{hopkins14}, which finds the mass-loading factor to scale as \mstarp$^{-0.35}$ when they calculate \mout at 0.25~R$_\mathrm{vir}$ \citep{muratov}, as shown by the dashed line in \autoref{fig:moutann}. The scaling of this relation is statistically similar to \autoref{eq:eta}, but the normalization is 4.6 times larger. However, we caution that the value of \mout from the simulations depends on the radius used to calculate \moutp, with \mout varying by a factor of 4 whether it is calculated at R$_\mathrm{vir}$ or at $0.25$R$_\mathrm{vir}$ \citep{muratov}. The discrepancy between our observations and the simulations does not necessarily mean that the simulation drives outflows that are too large: if other phases like the molecular gas, \ovi coronal gas, or hot X-ray-emitting plasma substantially contribute to the mass and momentum of the outflow, than our observed outflows do not account for the entire mass outflow. Future simulations that track the phase structure of outflows (e.g. molecular, photoionzied, transitional, and hot gas) can compare the mass-loading factors in different ionization phases to determine if the normalization discrepancy lies in unobserved phases or in the assumptions of the simulations.

\subsection{The energetics of driving galactic outflows}
\label{driving}
Driving galactic outflows in simulations is challenging. Cosmological simulations need to account for entire galaxies with physics on tens of kpc scales as well as detailed physics on sub-pc scales. In particular, resolving the size of a supernova blastwave is crucial to account for the energy and momentum of outflows because most of the supernova energy is radiated away in these small, dense regions when they are under-resolved \citep{katz92}. The lack of resolution prompted simulations to use computational methods, like temporarily turning off cooling immediately after a supernova or converting all of the supernova energy into kinetic energy \citep{navarro93, joki16}, to eliminate overcooling. More recently, simulations scale the outflow velocities and mass-loading factors to parameters of the host galaxies \citep{benson03, oppenheimer08, vogelsberger13, somerville15}, enabling simulations to generate outflows while remaining computationally feasible.

Driving outflows using scaling relations has a logical theoretical argument: star formation transfers momentum or energy into  the surrounding gas which accelerates the gas out of the galaxy. If observations can relate the transfer of momentum from star formation to the outflow, then simulations will drive realistic outflows using moderate computing resources. Typically, star formation is assumed to drive outflows by imparting momentum to the gas as
\begin{equation}
    \dot{p}_\mathrm{o} = \dot{M}_\mathrm{o} \mathrm{v}_\mathrm{o} = \zeta 
    \dot{p}_\mathrm{SFR}
\label{eq:mom}
\end{equation}
where v$_\mathrm{o}$ is the outflow velocity and $\zeta$ is the fraction of the total momentum produced by star formation ($\dot{p}_\mathrm{SFR}$) that is transferred to the outflow. 

Our sample has large $\zeta$ values: the median momentum of the outflow is 68\% of the momentum that is directly injected by supernovae. Moreover, outflows from galaxies below log(\mstarp) of 9 have more momentum than provided by supernovae alone. Stellar winds, the Sedov Taylor phase, and radiation pressure also add momentum to the gas, where {\small STARBURST99} models imply that radiation pressure and stellar winds add at least as much momentum as supernovae \citep{claus99}. Additionally, \autoref{eq:psfr} may under-estimate the total amount of momentum injected by supernovae by up to a factor of 10 during the Sedov-Taylor phase \citep{hopkins14, kim15, kim17}. These phases are crucial for understanding the amount of momentum transferred to the outflow. 

Regardless, \poutp/\psfr values greater than one require more sources of momentum than just the direct momentum from supernovae to produce the observed outflow momenta  \citep{hopkins12b}. The roles of different momentum sources are explored in \ion{H}{ii} regions of the Large and Small Magellanic Clouds where observations suggest that pressure from warm gas dominates the energetics, but there are also significant contributions from hot gas pressure and radiation pressure from dust at small radii \citep{lopez11, lopez14}. At a radius of 75~pc, \citet{lopez14} find that the dominant pressure source changes from radiation pressure to the thermal pressure of warm gas. The lower-mass galaxies tend to have small R$_\mathrm{i}$ (\autoref{tab:sample_out}), which may indicate that \psfr is different for these low-mass galaxies. Further, \autoref{eq:psfr} assumes that the observed star formation drives the observed outflows. The timescales defined by the velocity law imply that the gas is accelerated to terminal velocities in $1-10$~Myr (\autoref{eq:beta}), which is shorter than the $100$~Myr timescales of the IR and UV SFRs. Therefore, bursty star formation may impart different \pout than assumed in \autoref{eq:psfr}.

Theory typically solves \autoref{eq:mom} by assuming that v$_\mathrm{o}$ scales linearly with \vcircp, that $\dot{p}_\mathrm{SFR}$ is proportional to the SFR (\autoref{eq:psfr}), and that $\zeta$ is constant \citep{murray05}. Placing these assumptions into \autoref{eq:mom} and solving for \moutp/SFR gives a scaling of v$_\mathrm{circ}^{-1}$, which is often called a momentum driven outflow \citep{murray05}. A similar argument is made using the energy from star formation to derive the scaling of the mass-loading for an energy driven outflow as v$_\mathrm{circ}^{-2}$. Simulations often use these relations to produce galactic outflows (\autoref{comp}).

\begin{figure}
\includegraphics[width = 0.5\textwidth]{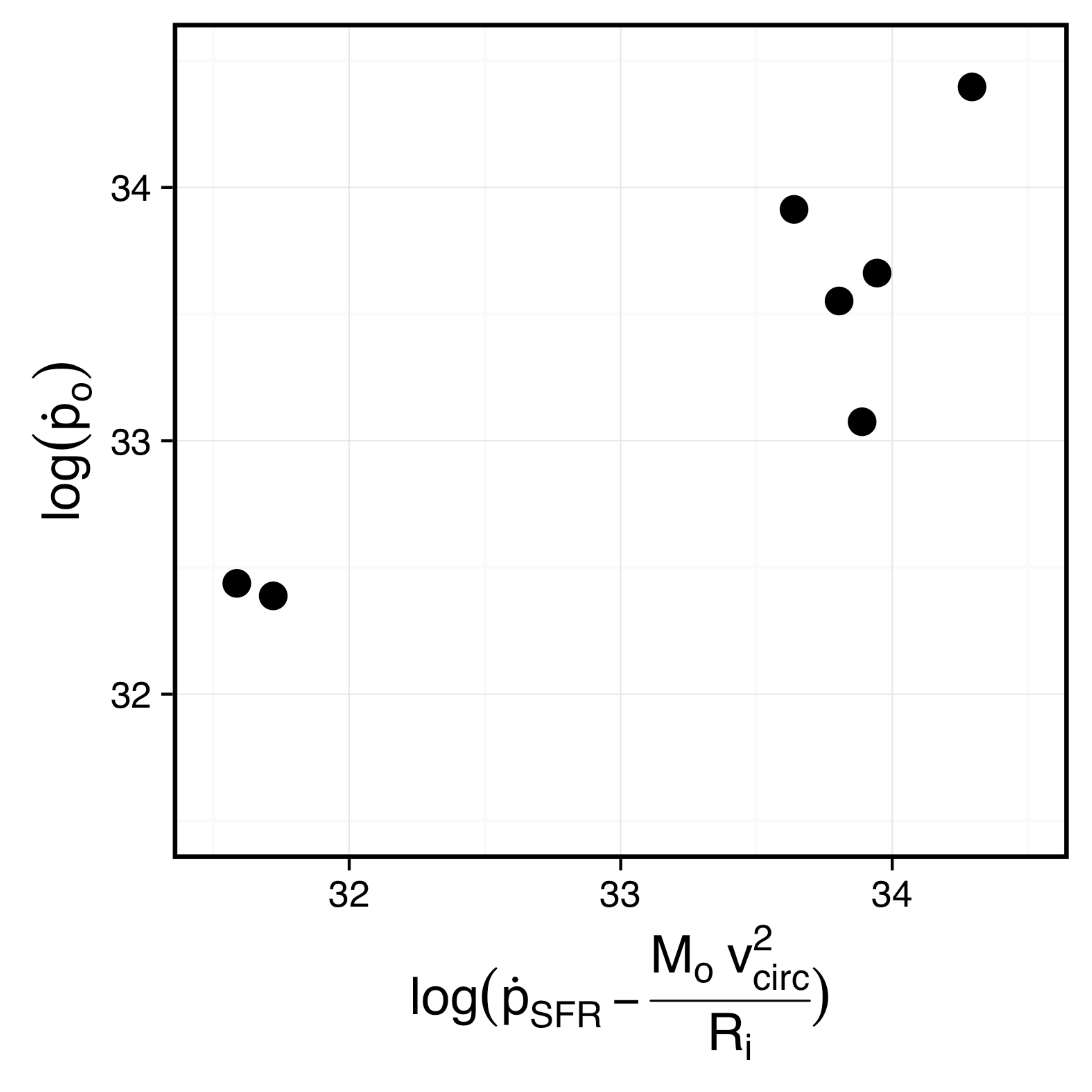}
\caption{A simplified model describing the relationship between the outflow momentum ($\dot{p}_\mathrm{out}$; in units of g~cm~s$^{-2}$) and the net force acting on the outflow (see \autoref{eq:drive}; in units of g~cm~s$^{-2}$). This simple model uses the observed star formation momentum to accelerate the outflow while gravity counteracts this acceleration to describe the observed outflow momenta.}
\label{fig:drive}
\end{figure}

However, our observations indicate that the situation is more nuanced because the momentum efficiency is not constant with \mstarp. The normalized momentum flux of SBS~1415+437, the lowest mass galaxy, is 20 times larger than the highest mass galaxy, NGC~6090. Furthermore, the scaling of the momentum efficiency with \mstar (\autoref{eq:pout}) indicates that momentum is either more easily dissipated in high-mass galaxies, or star formation injects more momentum in low-mass galaxies.

Gravitational drag could dissipate momentum. For example, the inward momentum deposition due to gravity ($\dot{p}_\mathrm{g}$) is equal to the retarding force of gravity (F$_\mathrm{g}$). If we assume a spherical mass distribution, the net force on the outflow is given as
\begin{equation}
 \dot{p}_\mathrm{o} = \dot{p}_\mathrm{SFR} - \frac{M_\mathrm{o} \mathrm{v}_\mathrm{circ}^2}{\mathrm{R}_\mathrm{i}}
 \label{eq:drive}
\end{equation}
Assuming that the bulk of outflowing mass (M$_\mathrm{o}$) is at R$_\mathrm{i}$ \citepalias{chisholm16b}, we use the R$_\mathrm{i}$ and M$_\mathrm{o}$ values from \autoref{tab:sample_out} to test whether this simplified model roughly reproduces the observed outflow momenta (\autoref{fig:drive}). This model does have a few limitations. First, the R$_\mathrm{i}$ values are sufficiently small such that the enclosed regions might not sample the full dark matter profile, and the gravitational force may not be accurately modeled by \vcircp. The model also does not account for the fact that low-mass galaxies have larger \pout than \psfrp, implying that supernovae are not the only momentum source accelerating these outflows (see the discussion above). This fact is partially offset because low-mass galaxies have lower covering fractions, possibly reducing the total amount of \psfr transferred to the outflow. Regardless, this simplified model has a relationship that is significant at the 3$\sigma$ significance, consistent with a unity slope, an R$^2$ of 0.85 and a Kendall's $\tau$ coefficient of 0.89. A simple model where gravity dissipates momentum from the outflow may explain the observed decrease in \poutp/\psfr with increasing \mstarp.

The energy efficiencies (\eoutp/\esfrp) of the outflows range between 0.9-21\%, implying that most of the energy from supernovae is dissipated by gravity, radiated away, or not in the photoionized phase. The \eoutp/\esfr values are consistent with the 1-10\% often found in numerical simulations \citep{thornton, efstathiou}. Importantly,  \eoutp/\esfr of the photoionized outflows increases with decreasing \mstar such that low-mass galaxies drive more efficient galactic outflows than high-mass galaxies. The combination of more efficient outflows and shallower potentials means that low-mass galaxies more efficiently remove gas from their star-forming regions \citep{dekel86}. 

Mass-loading factors above one indicate that the outflows deplete the gas within the galaxy more than the star formation does. \autoref{eq:eta} implies that the mass-loading factor exceeds one when log(\mstarp) is less than 9.7.  Outflows regulate the gas depletion of low-mass galaxies; star formation regulates the gas depletion of high-mass galaxies. This critical stellar mass is similar to the stellar mass found in \citetalias{chisholm15}, below which outflow velocities are faster than escape velocities. In fact, only NGC~7714 and NGC~6090, the two highest mass galaxies in the sample, do not have \siiv absorption at velocities greater than three times their \vcircp, a typical estimate of the escape velocity \citep{heckman2000}. Since the outflow is accelerated radially, the model presented here defines whether outflows escape the gravitational potential differently than previous models. In fact, with a radially accelerated outflow, each velocity interval is a snapshot of the outflow in time (or radius, or velocity) which may be accelerated to higher velocities at later times. Unfortunately, the density also declines rapidly with radius, making it impossible to observe whether the outflows from the highest-mass galaxies reach the escape velocity.

Galaxies in this sample with log(\mstarp) less than 9.7 have outflows that deplete more gas than their star formation does at velocities high enough to completely remove the gas from the galaxies. This may produce the bursty star formation histories of dwarf galaxies \citep{mateo} by removing most of the gas in a single burst of star formation \citep{dekel86}. Conversely, high-mass galaxies retain their outflows, and the gas reaccretes onto the galaxy as a galactic fountain \citep{shapiro}, providing a secondary source of star-forming material. Consequently, galaxies must be massive to retain outflowing gas, to efficiently convert gas into stars, and to have relatively constant star formation histories. Since star formation dominates the gas depletion in galaxies with halo masses greater than 10$^{11.4}$~M$_\odot$ \citep{moster10}, these galaxies retain a higher fraction of their total baryons as stars (i.e. they have a higher \mstarp/M$_\mathrm{halo}$ ratio). The observed \mstarp/M$_\mathrm{halo}$ relation peaks near 10$^{12}$~M$_\odot$ \citep{moster10} and declines at higher masses as AGN feedback becomes important, or as the halo becomes massive enough to shock heat accreting gas to high temperatures \citep[so called hot-mode accretion;][]{keres09}. Outflows are a significant component of galaxy evolution by shaping their star formation histories, regulating their gas content, and removing their baryons.

\subsection{The curious case of IRAS~08339+6517}
\label{iras}

A curious outlier to the above trends is the high-mass merging system IRAS~08339+6517. This galaxy has an anomalously low mass outflow rate, but extremely high outflow velocity. In \citetalias{chisholm15} we define a group of outflows with maximum velocities greater than 750~\kmsp, that have outflow velocities 32\% higher than other galaxies at similar stellar mass and SFR. IRAS~08339+6517 is the only galaxy in this sample that is in that high-velocity group. What makes IRAS~08339+6517 such a strange galaxy?

The power of the detailed profile fitting presented here is that we can differentiate groups of outflows based on the model fits to the absorption profile. The line profile of IRAS~08339+6517 does not fit the prescription in \autoref{profile} largely because C$_f$ does not vary coherently with velocity, as prescribed by the power-law scaling (see \autoref{fig:iras}). This is in sharp contrast to the rest of the sample (see \autoref{appendix}), which have similar C$_f$ distributions with a  C$_f$ power-law exponent ($\gamma$) of -0.88. To derive an upper limit to \mout for IRAS~08339+6517, we set the C$_f$ distribution by-hand to match the observations, with $\gamma = 0$ and C$_f$(R$_\mathrm{i}$)~=~1.

In \citetalias{chisholm16b} we use the observed C$_f$ power-law scaling of r$^{-0.8}$ to approximate the outflow as an ensemble of adiabatically expanding clouds in an adiabatically expanding medium. No variation in C$_f$ with velocity requires that the outflowing clouds expand at the same rate as geometric dilution. Alternatively, the high-velocity outflows might not fit the physical picture presented in \autoref{profile}. One possibility is that the absorption does not correspond to outflowing gas, rather tidal interactions have distributed gas at a wide range of velocities along the line-of-sight, creating the unity covering of the source in \autoref{fig:iras}. 

NGC~7552, another merger with a high-velocity outflow (\siii velocity of -1043~\kmsp), has a similarly flat \siiv C$_f$ distribution. Unfortunately, strong geocoronal lines contaminate the \oi and \siii~1304~\AA\ profile, making photoionization models impossible for this galaxy. While two galaxies do not constitute a complete group, their similarities suggest that the highest velocity outflows have different line profiles, and including different types of outflow profiles may increase the scatter,  increase measured velocities, and confuse the derivation of trends between outflow properties and host galaxy properties. 

\section{CONCLUSION}
Here we calculate the mass (\moutp), energy (\eoutp), and momentum (\poutp) outflow rates for a sample of 7 nearby star-forming galaxies. We use a Sobolev approximation and detailed photoionization models to determine the quantities with fewer assumed parameters than previous studies. These observations describe how efficiently photoionized outflows remove mass and momentum from star-forming regions. For example, galaxies in the sample with log(\mstarp) less than 9.7 eject more mass in their outflow than they form into stars, at velocities that exceed their escape velocities. The momentum of outflows from low-mass galaxies is greater than the momentum directly injected from supernovae alone, implying that there must be additional momentum sources driving the outflows. Only 1-20\% of the energy released by supernovae is converted into the kinetic energy of the photoionized outflow, the rest is dissipated by gravity, radiated away, or in a different temperature phase. The values of the mass-loading factor, \poutp/\psfrp, and \eoutp/\esfr describe how efficiently outflows remove gas from galaxies, and demonstrate that the evolution of the gas content of low-mass galaxies is dominated by galactic outflows. 

We find a 3$\sigma$ relation between the galactic stellar mass and the mass-loading factor (\moutp/SFR; \autoref{eq:mout} and \autoref{fig:mout}). The \poutp/\psfr ratio is also correlated at the 2.5$\sigma$ significance with \mstar (\autoref{fig:pout}). The momenta are described by a simple model where star formation drives the gas outward while gravity counteracts the acceleration (\autoref{eq:drive} and \autoref{fig:drive}). Additionally, low-mass galaxies are more energy efficient than high-mass galaxies, suggesting that dwarfs efficiently remove gas from their star-forming regions. The mass outflow rates presented here are five times weaker than simulations typically implement, but have similar scalings. This normalization discrepancy is likely because we only observe the photoionized outflow and there is a substantial amount of mass in other outflowing phases.

In a future paper, we will discuss how the outflow metallicity, inner radius, and density profile changes with star formation rate and stellar mass. These results are crucial to understanding the enrichment of the circum-galactic medium, and how outflows transport metals out of star-forming regions to establish the mass-metallicity relationship.

\section*{Acknowledgments}
We thanks the anonymous referee that provided comments and suggestions that significantly improved the manuscript. Joseph Cassinelli inspired this work with helpful conversations and notes.
We thank Bart Wakker for help with the data reduction.

Support for program 13239 was provided by NASA through a grant from the Space Telescope Science Institute, which is operated by the Association of Universities for Research in Astronomy, Inc., under NASA contract NAS 5-26555. All of the data presented in this paper were obtained from the Mikulski Archive for Space Telescopes (MAST). STScI is operated by the Association of Universities for Research in Astronomy, Inc., under NASA contract NAS 5-26555. Support for MAST for non-HST data is provided by the NASA Office of Space Science via grant NNX09AF08G and by other grants and contracts. 

\bibliographystyle{mnras}
\bibliography{outflow}

\bsp	
\label{lastpage}

\appendix
\section{Optical Depth and Covering Fraction Plots}
\label{appendix}
Here we include the plots of the velocity-resolved $\tau$ (upper panels) and C$_f$ (lower panels) for the entire sample. Fits to the lines, using \autoref{eq:cfbeta} are shown by the solid line. The velocity resolution is marked in the upper portion of each panel. 
\begin{figure*}
\includegraphics[width = \textwidth]{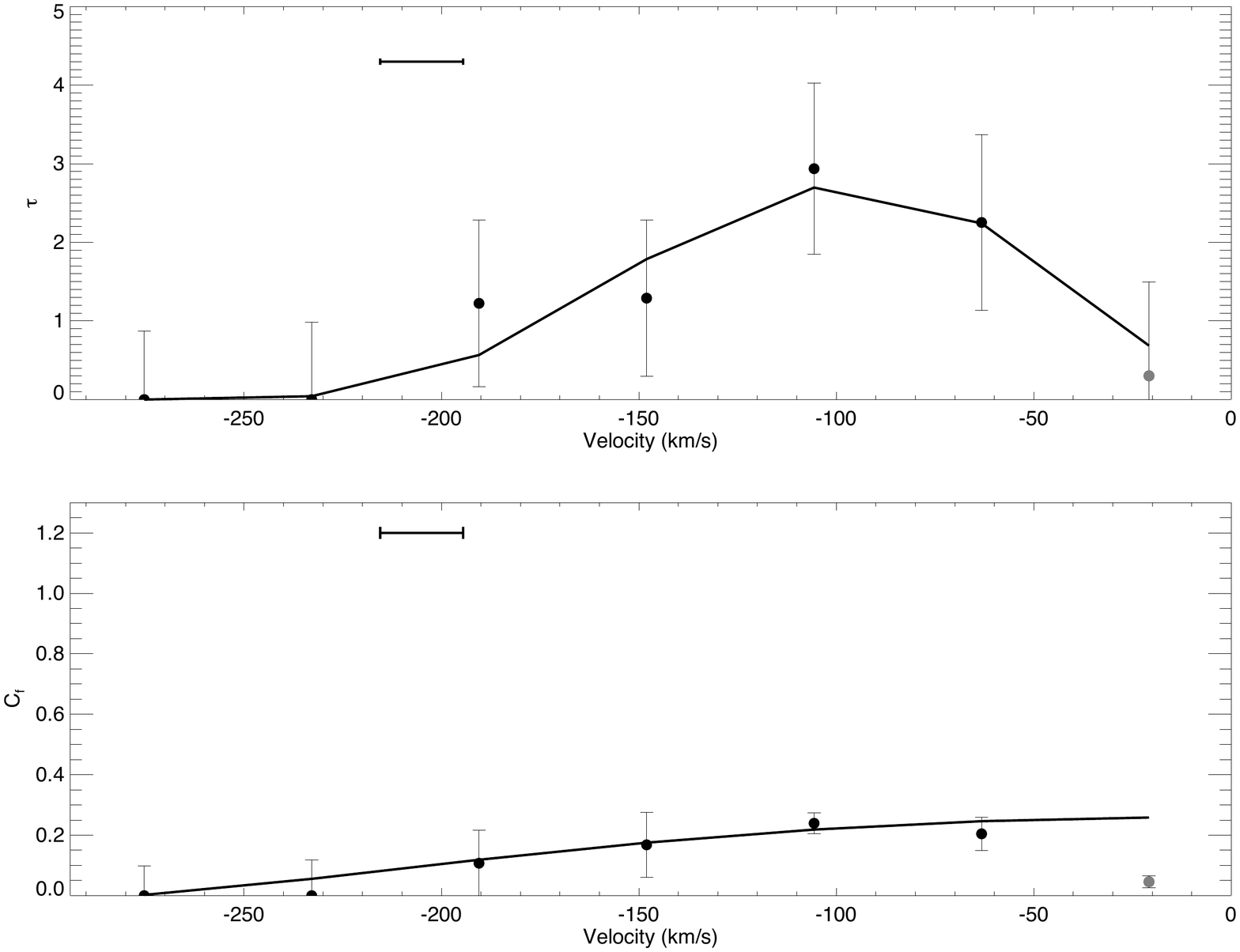}
\caption{The velocity resolved \siiv optical depth  ($\tau$; upper panel) and covering fraction (C$_f$; lower panel) for SBS1415+437, as derived from \autoref{eq:cf}. The $\tau$ is fit (solid line) assuming a Sobolev optical depth, and the C$_f$ is fit assuming a radial power-law C$_f$, as given by \autoref{eq:cfbeta}. Gray points are excluded from the fit due to contamination of resonance emission. The spectral resolution is given as a bar in the upper portion of each panel.}
\label{fig:sbs}
\end{figure*}
\begin{figure*}
\includegraphics[width = \textwidth]{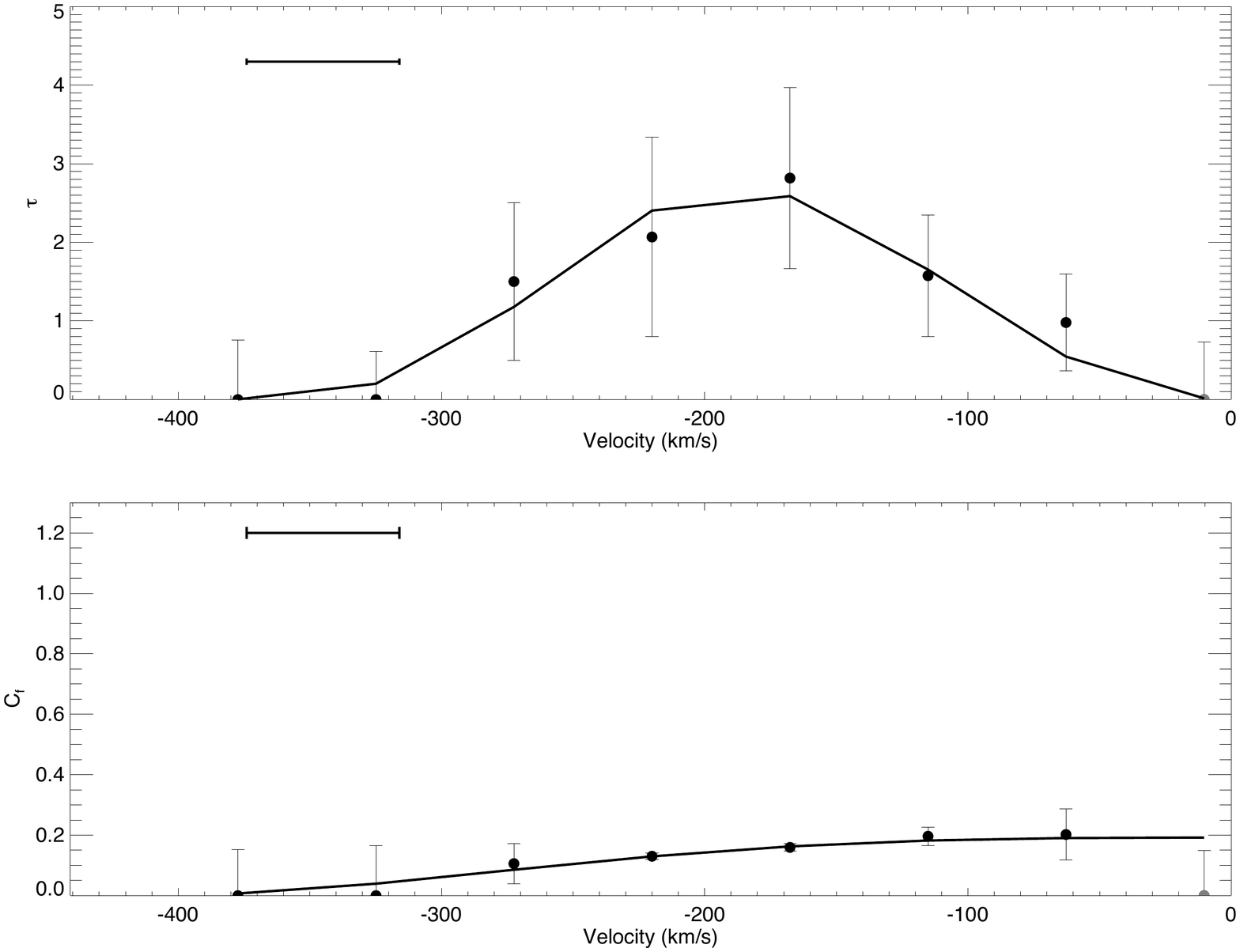}
\caption{Same as \autoref{fig:sbs} but for 1~Zw~18. .}
\label{fig:1zw18}
\end{figure*}
\begin{figure*}
\includegraphics[width = \textwidth]{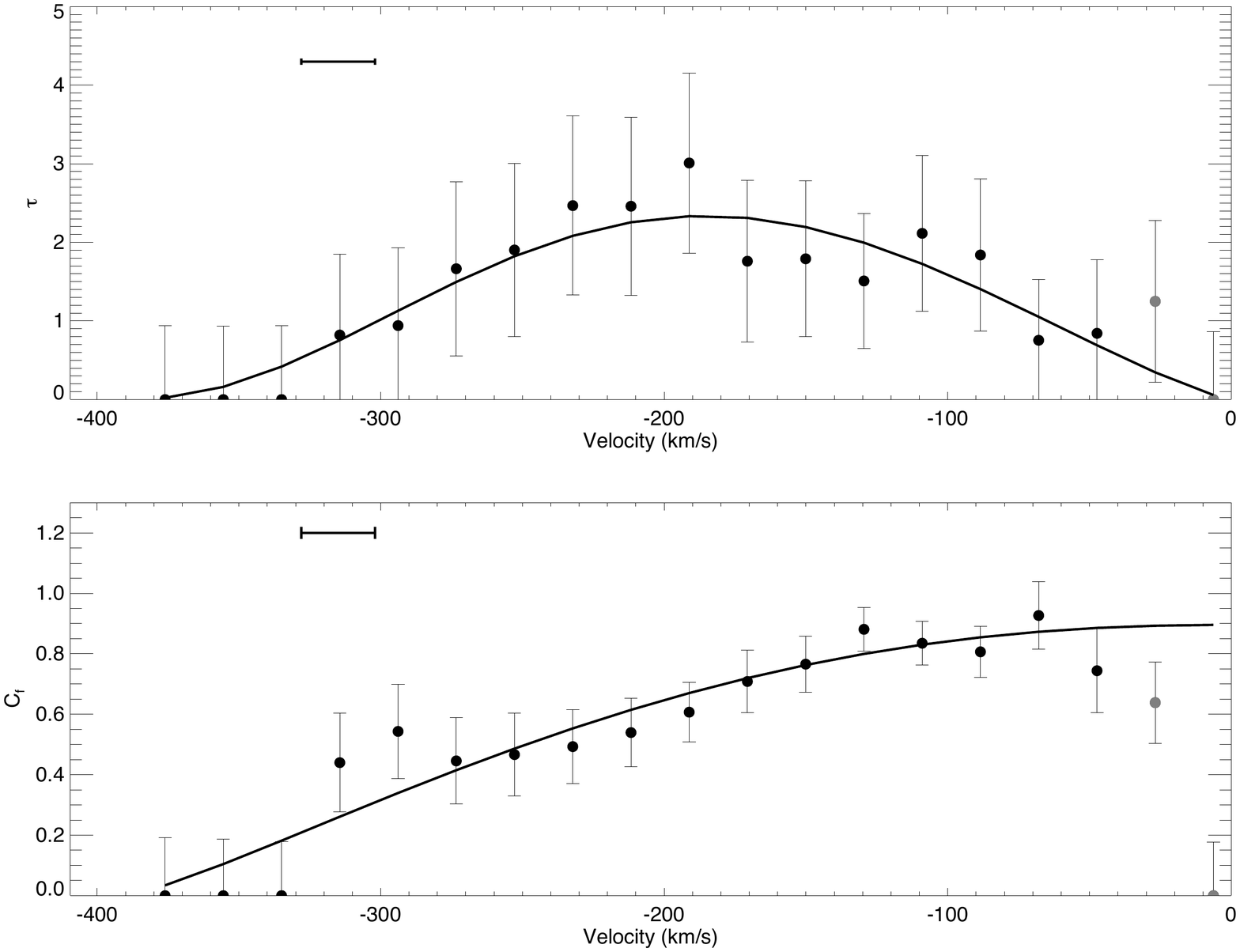}
\caption{Same as \autoref{fig:sbs} but for MRK~1486. }
\label{fig:mrk1486}
\end{figure*}
\begin{figure*}
\includegraphics[width = \textwidth]{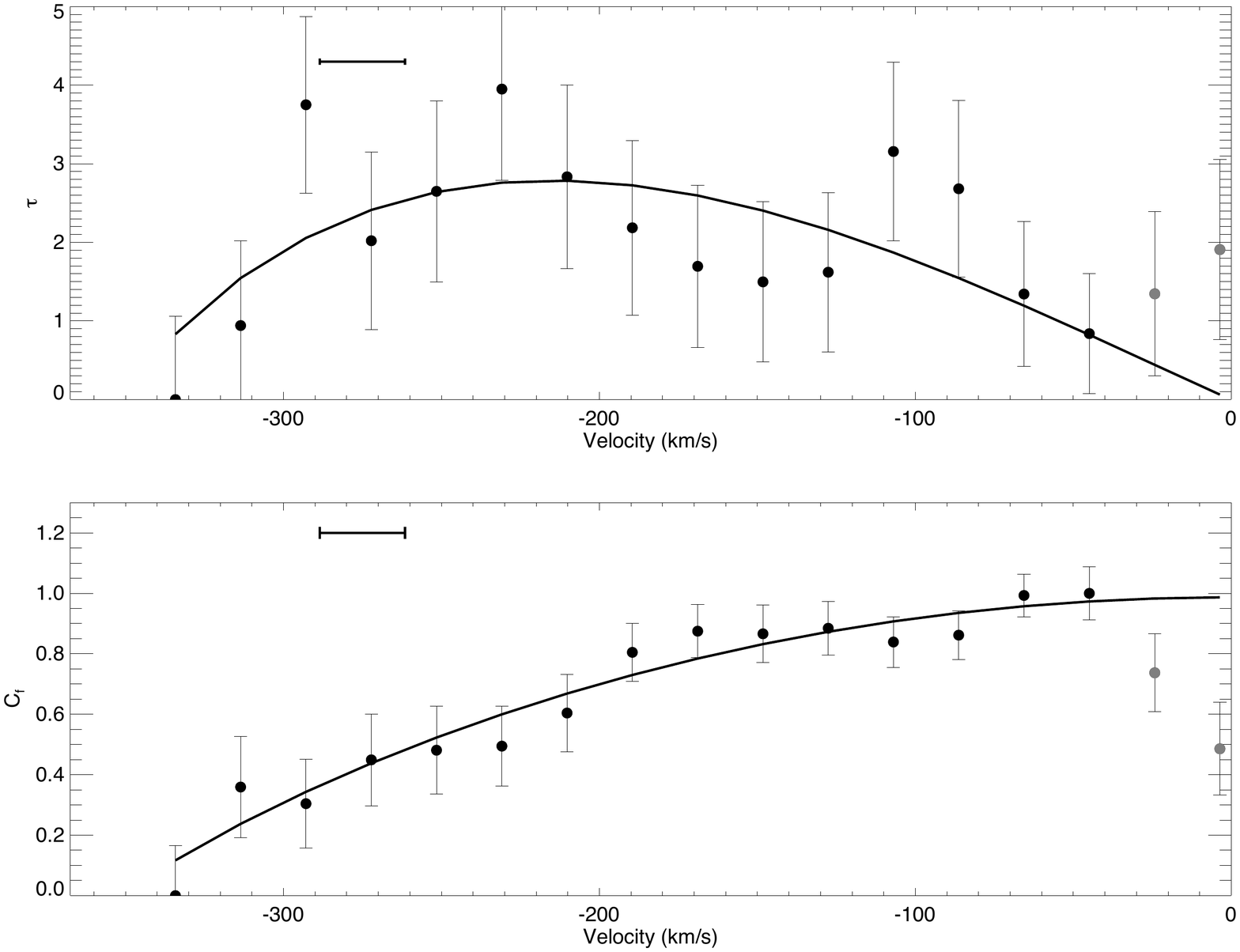}
\caption{Same as \autoref{fig:sbs} but for KISSR~1578.}
\label{fig:kissr1578}
\end{figure*}
\begin{figure*}
\includegraphics[width = \textwidth]{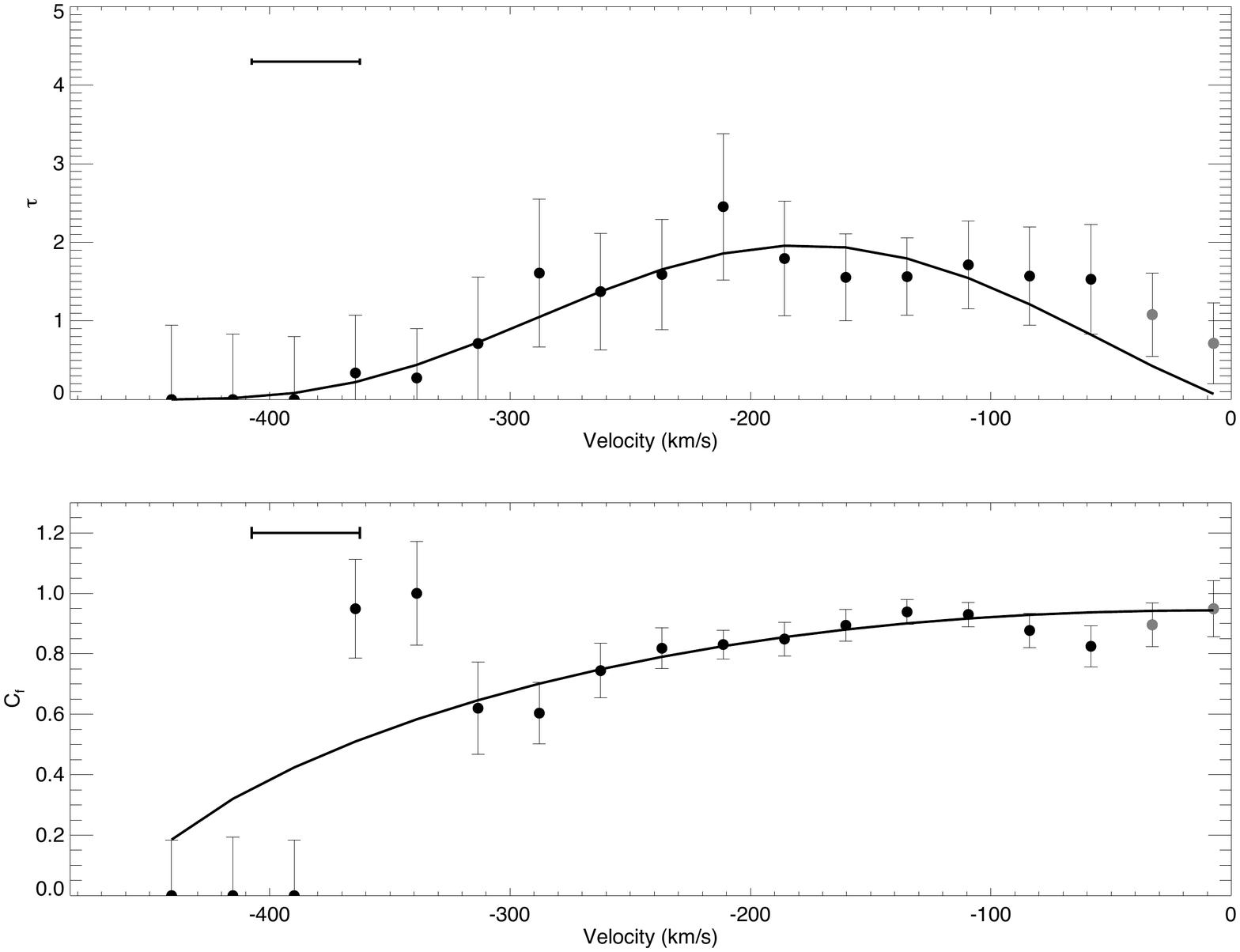}
\caption{Same as \autoref{fig:sbs} but for Haro~11.}
\label{fig:haro11}
\end{figure*}
\begin{figure*}
\includegraphics[width = \textwidth]{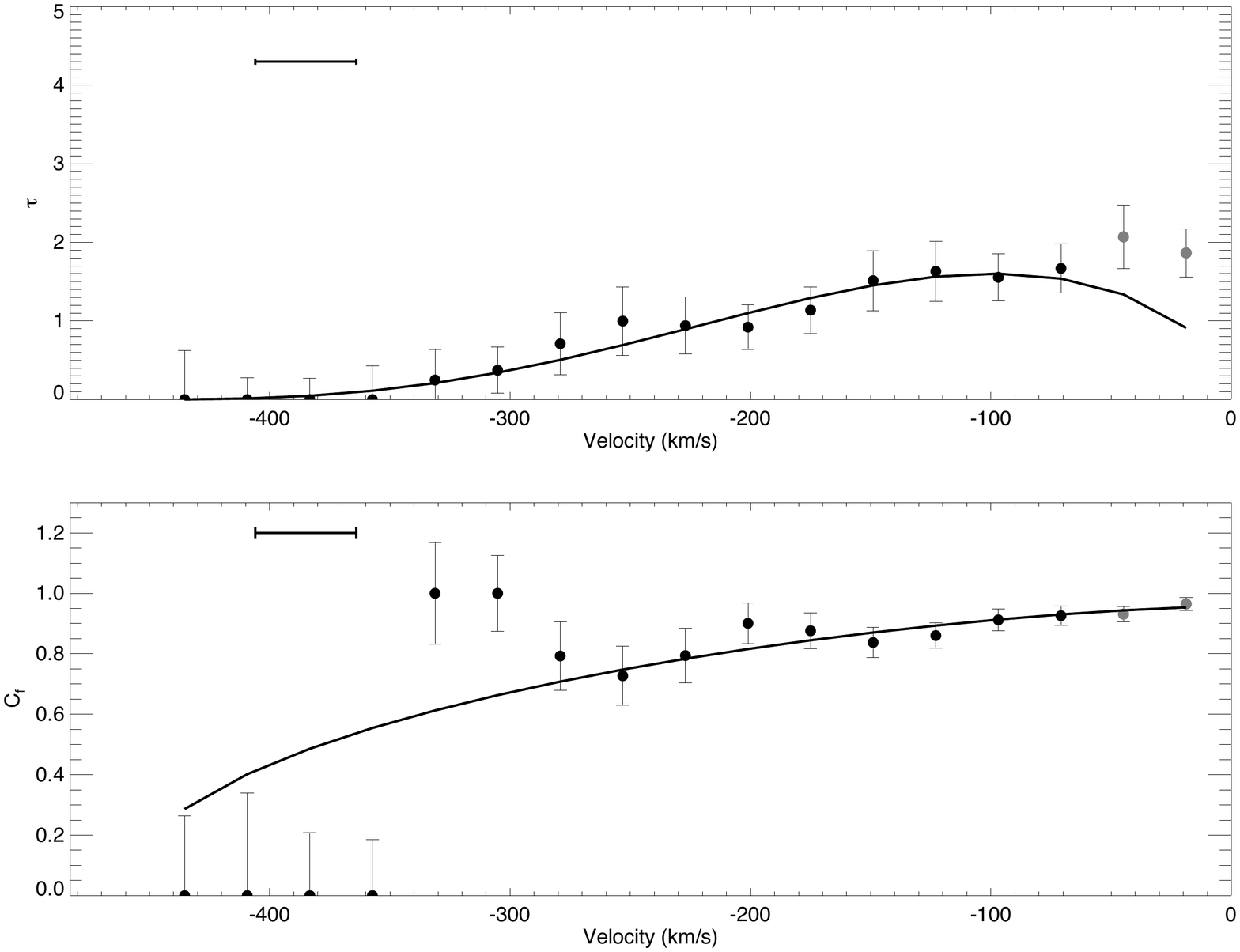}
\caption{Same as \autoref{fig:sbs} but for NGC~7714.}
\label{fig:ngc7714}
\end{figure*}
\begin{figure*}
\includegraphics[width = \textwidth]{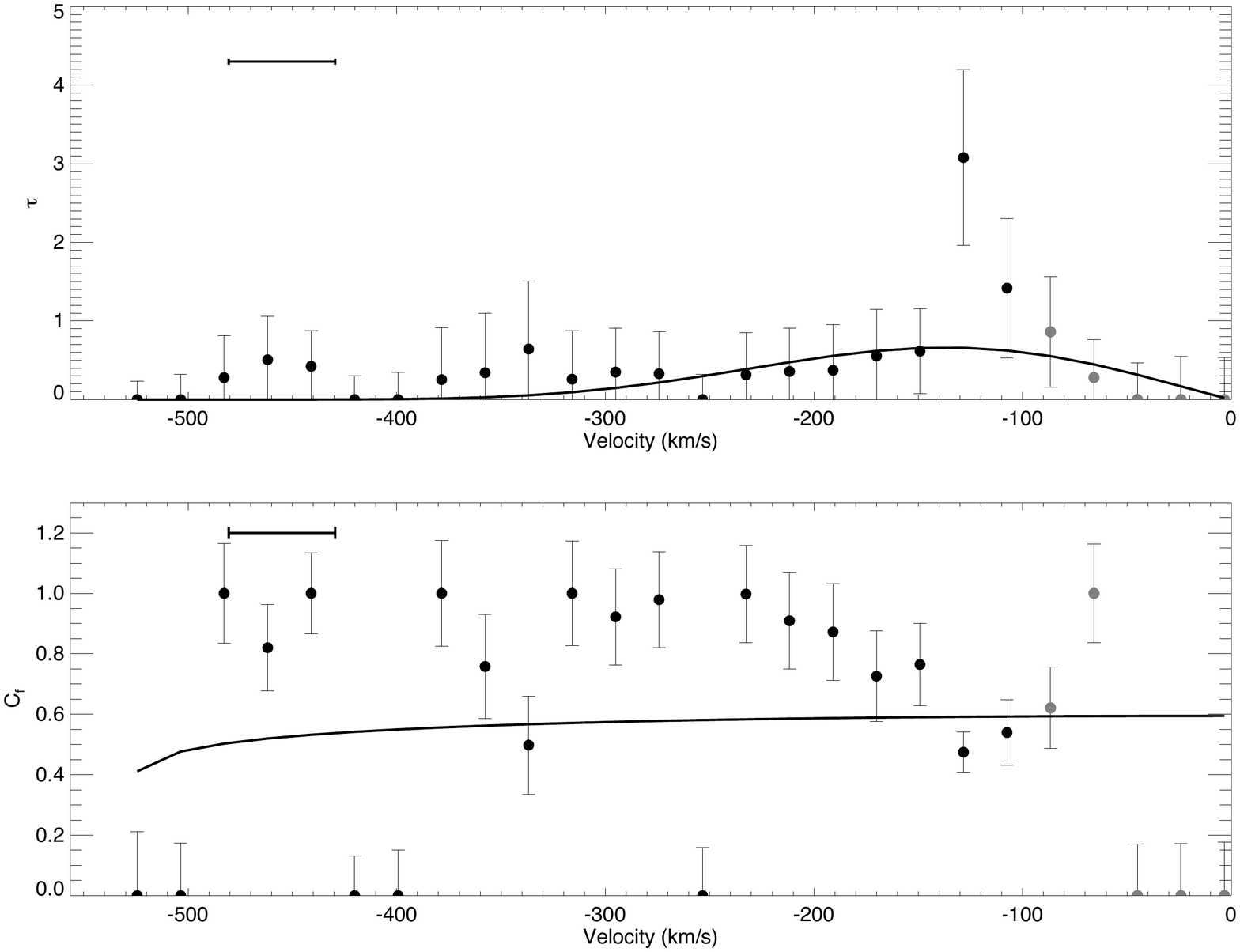}
\caption{Same as \autoref{fig:sbs} but for IRAS08449+6517. This galaxy is not included in the analysis because the line profile cannot match the proposed model of \autoref{profile}}
\label{fig:iras}
\end{figure*}
\begin{figure*}
\includegraphics[width = \textwidth]{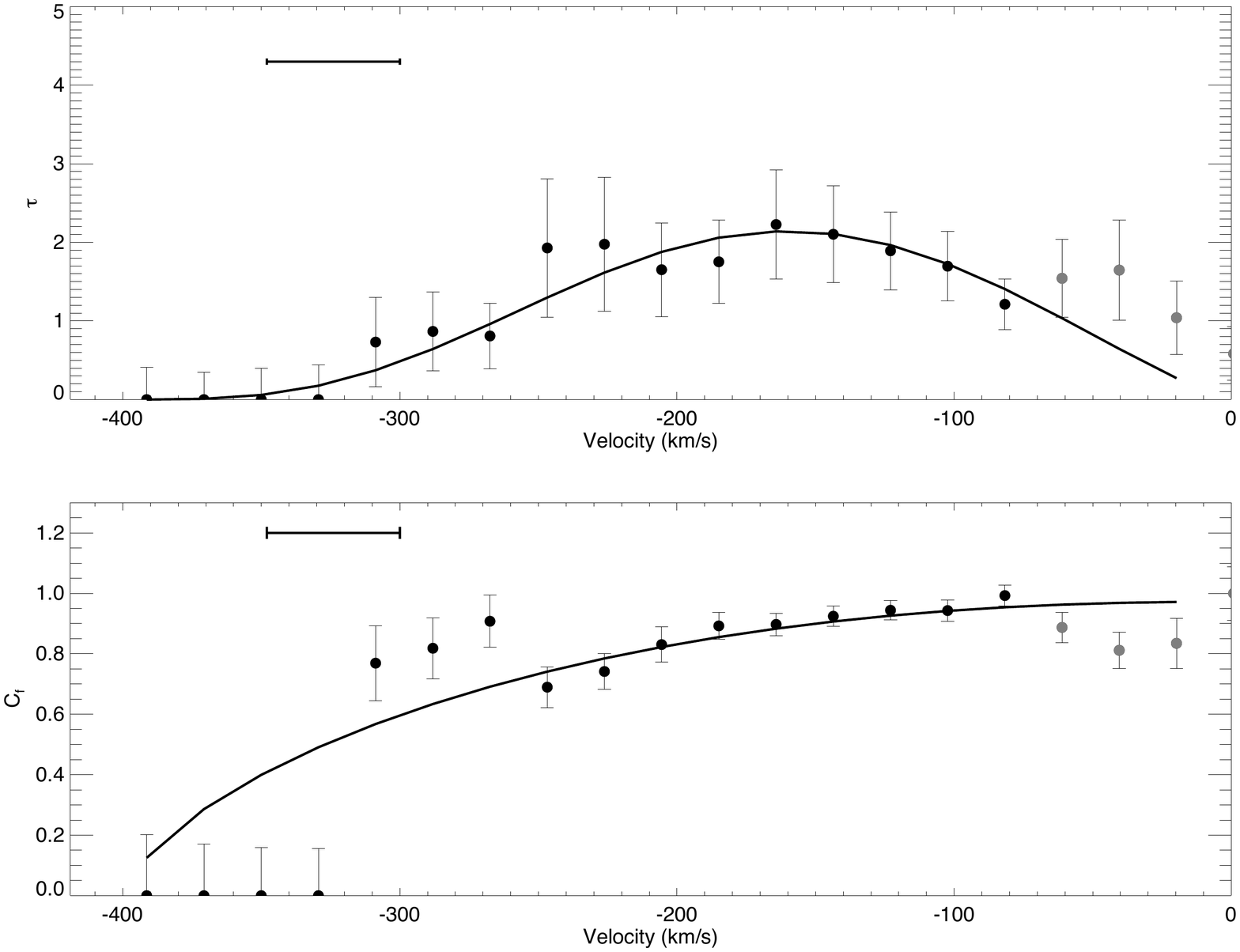}
\caption{Same as \autoref{fig:sbs} but for NGC~6090. }
\label{fig:NGC6090}
\end{figure*}

\end{document}